\documentclass[twocolumn,tighten]{aastex63}
\usepackage{apjfonts}
\usepackage{graphicx}
\usepackage{xcolor}
\newcommand{\secpoint}{\mbox{$''\mskip-7.6mu.\,$}}

\definecolor{khpink}{rgb}{0.858, 0.188, 0.478}

\definecolor{RHcomment}{rgb}{0.427, 0.714, 1}

\received{11/25/19}
\revised{2/28/20}
\accepted{3/7/20}

\begin{document}

\title{Simulating \textit{JWST}/NIRCam Color Selection of High-Redshift Galaxies}

\shorttitle{NIRCam High-Redshift Dropout Galaxies with JAGUAR}
\shortauthors{HAINLINE ET AL.}

\author[0000-0003-4565-8239]{\sc Kevin N. Hainline}
\affiliation{Steward Observatory, University of Arizona, 933 North Cherry Avenue, Tucson, AZ 85721, USA}

\author[0000-0002-4684-9005]{Raphael E. Hviding}
\affiliation{Steward Observatory, University of Arizona, 933 North Cherry Avenue, Tucson, AZ 85721, USA}

\author[0000-0002-7893-6170]{Marcia Rieke}
\affiliation{Steward Observatory, University of Arizona, 933 North Cherry Avenue, Tucson, AZ 85721, USA}

\author[0000-0003-4702-7561]{Irene Shivaei}
\affiliation{Hubble Fellow, Steward Observatory, University of Arizona, 933 North Cherry Avenue, Tucson, AZ 85721, USA}

\author[0000-0003-4564-2771]{Ryan Endsley}
\affiliation{Steward Observatory, University of Arizona, 933 North Cherry Avenue, Tucson, AZ 85721, USA}

\author{Emma Curtis-Lake}
\affiliation{Kavli Institute for Cosmology, Madingley Road, Cambridge CB3 0HA, UK}
\affiliation{Cavendish Laboratory, University of Cambridge, 19 JJ Thomson Avenue, Cambridge CB3 0HE, UK}

\author[0000-0001-8034-7802]{Renske Smit}
\affiliation{Cavendish Laboratory, University of Cambridge, 19 JJ Thomson Avenue, Cambridge CB3 0HE, UK}

\author[0000-0003-2919-7495]{Christina C. Williams}
\altaffiliation{NSF Fellow} 
\affiliation{Steward Observatory, University of Arizona, 933 North Cherry Avenue, Tucson, AZ 85721, USA}

\author[0000-0002-8909-8782]{Stacey Alberts}
\affiliation{Steward Observatory, University of Arizona, 933 North Cherry Avenue, Tucson, AZ 85721, USA}

\author{Kristan N K Boyett}
\affiliation{Department of Physics, University of Oxford, Denys Wilkinson Building, Keble Road, OX1 3RH, UK}

\author{Andrew J. Bunker}
\affiliation{Department of Physics, University of Oxford, Denys Wilkinson Building, Keble Road, OX1 3RH, UK}

\author{Eiichi Egami}
\affiliation{Steward Observatory, University of Arizona, 933 North Cherry Avenue, Tucson, AZ 85721, USA}

\author[0000-0003-0695-4414]{Michael V. Maseda}
\affiliation{Leiden Observatory, Leiden University, P.O. Box 9513, NL-2300 RA Leiden, the Netherlands}

\author[0000-0002-8224-4505]{Sandro Tacchella}
\affiliation{Center for Astrophysics, Harvard \& Smithsonian, 60 Garden Street, Cambridge, MA 02138, USA}

\author[0000-0001-9262-9997]{Christopher N.A. Willmer}
\affiliation{Steward Observatory, University of Arizona, 933 North Cherry Avenue, Tucson, AZ 85721, USA}

\begin{abstract}

The NIRCam instrument on the upcoming \textit{James Webb Space Telescope} (\textit{JWST}) will offer an unprecedented view of the most distant galaxies. In preparation for future deep NIRCam extragalactic surveys, it is crucial to understand the color selection of high-redshift galaxies using the Lyman dropout technique. To that end, we have used the JAdes extraGalactic Ultradeep Artificial Realizations (JAGUAR) mock catalog to simulate a series of extragalactic surveys with realistic noise estimates. This enables us to explore different color selections and their impact on the number density of recovered high-redshift galaxies and lower-redshift interlopers. We explore how survey depth, detection signal-to-noise ratio, color selection method, detection filter choice, and the presence of the Ly$\alpha$ emission line affects the resulting dropout selected samples. We find that redder selection colors reduce the number of recovered high-redshift galaxies, but the overall accuracy of the final sample is higher. In addition, we find that methods that utilize two or three color cuts have higher accuracy because of their ability to select against low-redshift quiescent and faint dusty interloper galaxies. We also explore the near-IR colors of brown dwarfs and demonstrate that, while they are predicted to have low on-sky densities, they are most likely to be recovered in F090W dropout selection, but there are color cuts which help to mitigate this contamination. Overall, our results provide NIRCam selection methods to aid in the creation of large, pure samples of ultra high-redshift galaxies from photometry alone.

\end{abstract}

\keywords{galaxies: distances and redshifts -- galaxies: high-redshift}

\section{Introduction}
\label{sec:intro}
The discovery and characterization of high-redshift ($z > 6$) galaxies offers fundamental insights into galaxy assembly and star formation, including the creation of dust and metals, in the first billion years of the history of the universe. Deep imaging with the Wide Field Camera 3 (WFC3) instrument on board the \textit{Hubble Space Telescope} (\textit{HST}) has revealed samples of galaxies at these redshifts \citep{bouwens2003, bouwens2004, bouwens2007, bouwens2008, bunker2004, bunker2010, mclure2010, wilkins2011, lorenzoni2011, lorenzoni2013}, including an intriguing, if limited, population of ultra-high redshift galaxies at $z > 10$ \citep{oesch2014, oesch2015, oesch2018, zitrin2014, infante2015, ishigaki2015, mcleod2016, salmon2018}. Assembling larger populations of galaxies at higher redshifts is challenging due to the lack of infrared coverage of the instruments on \textit{HST} (the longest wavelength filter on WFC3 is at 1.6 $\mu$m), the limited sensitivity and low resolution of observations made at longer wavelengths by the \textit{Spitzer Space Telescope}, and infrared atmospheric absorption for ground-based observations. Overcoming these limitations is fundamental for understanding the evolution of the earliest galaxies \citep[see reviews by][]{dunlop2013, stark2016}.

The selection of high-redshift galaxies is crucial for our understanding of reionization, where the neutral hydrogen that filled the universe after recombination was ionized in a process thought to be driven by early star-forming galaxies between $z \sim 6 - 10$ \citep{robertson2015}, although accretion onto supermassive black holes is also thought to be a contributing factor \citep{giallongo2015,madau2015,onoue2017}. By characterizing the galaxies that comprise the faint end of the UV luminosity function, the exact source and timescale of reionization can be understood. In addition, observations of these galaxies give us insight into the evolution of the star-formation rate density in the early universe, which has been observed to increase by almost an order of magnitude in the 170 million years between $8 < z <  10$ \citep{oesch2014, oesch2018, ishigaki2018}, although observations by \citet{mcleod2016} indicate a shallower evolution during this period. This tension may be due to cosmic variance and small sample sizes, providing a clear impetus to uncover larger samples of ultra-high-redshift galaxies.

A widely used method for selecting galaxies at high redshift involves searching for their redshifted Lyman break, a feature in their spectrum caused by the absorption of extreme ultraviolet radiation by neutral hydrogen in the intergalactic medium along the line of sight and surrounding a given galaxy. In this technique, a galaxy observed in a filter that probes a wavelength range bluewards of the Lyman break will have reduced flux compared to a filter that lies to the red of the break. As a result, by selecting for galaxies with extreme red colors in adjacent bands, a rough estimate of the redshift of the galaxy can be obtained \citep{guhathakurta199}. Galaxies selected in this way are referred to as ``dropouts.'' This method was used to assemble a large sample of galaxies at $z = 2 - 4$ using ground-based observations in the optical $U$, $G$, and $R$ filters, which was subsequently observed spectroscopically to confirm individual galaxy redshifts \citep{steidel1996, steidel1999, steidel2003}. This technique has subsequently been supported with spectroscopic observations of galaxies out to $z \sim 8$ \citep{bunker2003, stanway2004, vanzella2009, vanzella2011, stark2010, ono2012, schenker2012, shibuya2012, cassata2015, oesch2015b, robertsborsani2016, song2016, tasca2017}.  

    \begin{figure}[t!]
    \centering
    \includegraphics[width=\columnwidth]{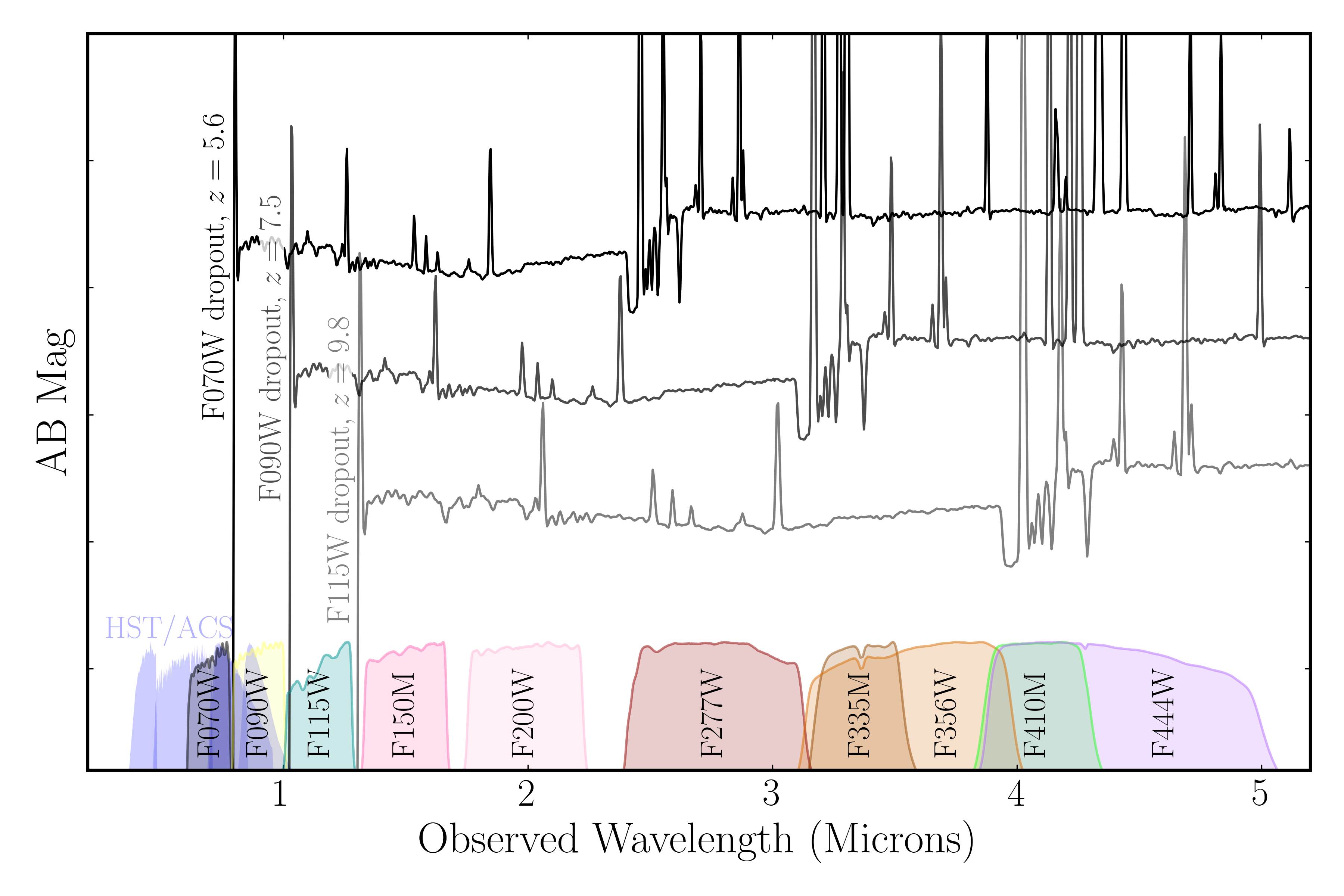}
    \caption{Example JAGUAR mock galaxy SEDs for F070W ($z = 5.6$, black), F090W ($z = 7.5$, grey), and F115W ($z = 9.8$, light grey) dropout galaxies. We also plot the HST/ACS bands we use in this work in light blue, and the \textit{JWST}/NIRCam filters in multiple colors as labelled. 
    \label{fig:LyADropoutExample} } 
    \end{figure}

An alternate method for estimating accurate photometric redshifts relies on modeling a galaxy's full spectral energy distribution (SED). The use of this method requires additional observed photometry over what is often needed for Lyman dropout selection, as well as a diverse suite of observed galaxy templates or stellar population synthesis models. In addition, it is less straightforward to understand the sample selection and survey completeness for SED modeling techniques than for color selection methods, and color selection is significantly quicker than full template fitting. For these reasons, in this paper we will focus on dropout selection of high-redshift galaxies. 

The near-infrared wavelength coverage of \textit{HST} and \textit{Spitzer} has been used to select dropout galaxies out to the current redshift frontier of $z = 9 - 11$ \citep{ellis2013, oesch2013, mclure2013, bouwens2015}. At higher redshifts, the Lyman break is shifted further into the infrared, and this technique is therefore limited by the lack of HST WFC3 filters at wavelengths longer than 1.6 $\mu$m. The infrared wavelength coverage and sensitivity of the Near Infrared Camera (NIRCam) instrument on the James Webb Space Telescope \citep[\textit{JWST},][]{gardner2006} will enable the discovery of galaxies out to $z > 15$. Following the projected launch of \textit{JWST} in 2021, NIRCam will provide 0.7 $\mu$m to 5 $\mu$m imaging over a 9.7 arcmin$^2$ field of view at resolutions of 0\secpoint04 - 0\secpoint1. NIRCam offers excellent sensitivity in this wavelength range, with 10$\sigma$ point source depths of 28 magnitude (AB) achievable in only 2 ksec at 2 $\mu$m. As \textit{JWST} is designed for only a nominal 5 - 10 year mission, it is imperative that we explore the ways in which NIRCam observations can be quickly and efficiently leveraged to assemble large samples of high-redshift galaxies.

To that end, in this study we use a catalog of mock galaxies to explore the relationship between various color selection methods and the properties of recovered high-redshift dropout galaxies. We use the JAdes extraGalactic Ultradeep Artificial Realizations (JAGUAR) mock catalog \citep{williams2018}, which was developed by members of the joint NIRCam and NIRSpec Guaranteed Time Observation (GTO) teams to aid in preparing for the early observations that will be made with \textit{JWST}, with a focus on the \textit{JWST} Deep Extragalactic Survey (JADES) GTO program. JAGUAR offers a catalog of photometry and spectra for mock galaxies along with self-consistent modeling of strong UV and optical emission lines, and was created using current observations of the number counts of galaxies as a function of UV luminosity and mass. To prepare for future deep \textit{JWST}/NIRCam surveys, we simulate NIRCam noise at various observational depths to explore how color cuts affect the number densities, redshift distributions, and intrinsic properties of recovered mock galaxies. We explore dropout selection using both \textit{JWST}/NIRCam filters alone as well as selection with NIRCam + \textit{HST}/ACS filters which are helpful for imaging below the Lyman break and rejecting low-redshift interlopers. The goal of this present study is not to provide canonical color cuts, but rather to demonstrate the types of color cut selection scenarios that can be employed to assemble galaxy samples at multiple redshift ranges. 

We begin by outlining the creation of our photometric catalogs with realistic noise properties in Section \ref{sec:methods}. There we provide an overview of JAGUAR, describe the methods by which we generate estimates of NIRCam noise, and discuss the overall design of the surveys we used to explore NIRCam color space. We outline our results in Section \ref{sec:results} for both simple single color selection and more complex, multi-color, selection. Additionally, we look at how dropout galaxy recovery is impacted by filter choice with an eye towards designing surveys that best utilize the unique dichroic beam splitter on board NIRCam. We also explore other common statistics used to separate interlopers, and the NIRCam colors of brown dwarfs. Finally, we discuss these results in Section \ref{sec:discussion}, and conclude in Section \ref{sec:conclusions}. Throughout we adopt a cosmology with H$_0 = 70$ km s$^{-1}$ Mpc$^{-1}$, $\Omega_M = 0.3$, and $\Omega_\Lambda = 0.7$. All magnitudes are presented in the AB system \citep{okegunn1983}. 

\section{Methods}
\label{sec:methods}

To explore the impact that color selection choices can make on recovered galaxy samples, we require a mock catalog that is diverse in star formation properties, redshifts, stellar masses, and dust attenuation as well as simulated observational noise at multiple depths. We also depend on statistical measures of how successful a given set of color cuts is at recovering high-redshift galaxy samples. In this section, we outline the JAGUAR catalog and describe our method for adding photometric noise to the JAGUAR photometry to produce mock observational catalogs at different simulated exposure times. We then describe the figures of merit we will use to compare the results from changing color selection methods, and finally, we discuss how we use these noisy photometric data to explore the NIRCam color space.

    \begin{figure*}[ht!]
    \centering
    \includegraphics[width=\textwidth]{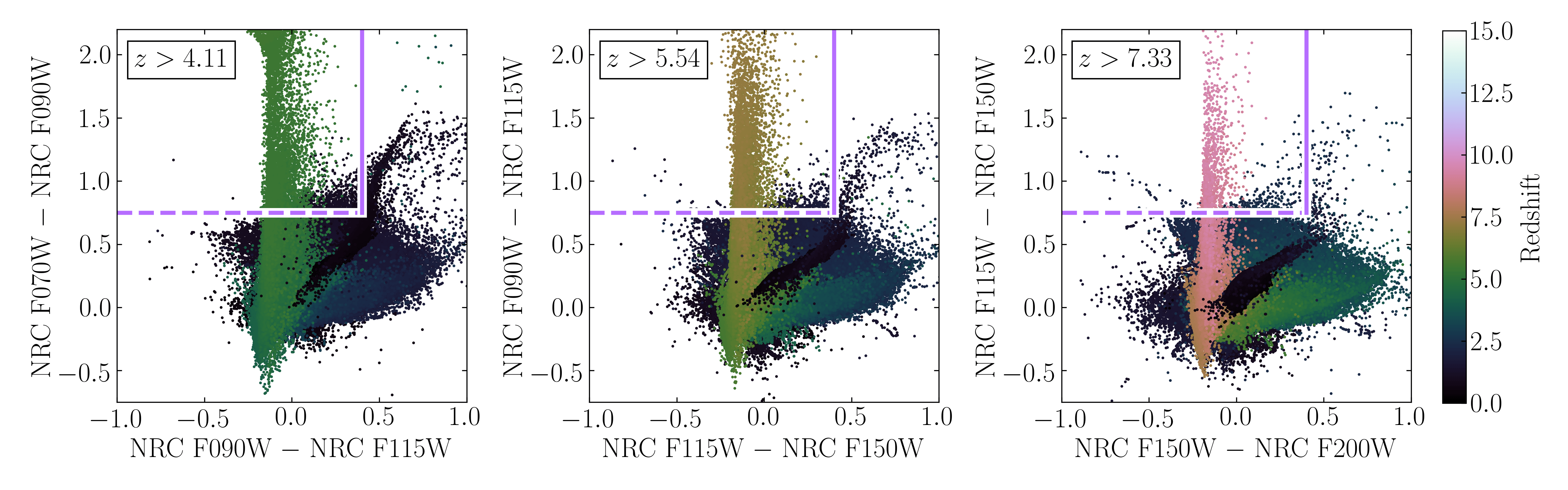}
    \caption{
        \label{fig:jaguarcolor} NIRCam color-color plots for a 10'x10' section of the JAGUAR catalog, with mock galaxies at $z = 0.2 - 15$, without adding noise, with points colored by catalog redshift values, as given by the colorbar on the right side of the figure. The left panel shows F070W dropouts at $z \sim 5.5$, the center panel shows F090W dropouts at $z \sim 7.3$, and the right panel shows F115W dropouts at $z \sim 9.7$. In addition, in each panel, populations of lower redshift mock galaxies have red colors on both axes, and most selection methods at these redshifts will deliberately exclude these objects. We plot an example two-color selection method in each panel in lavender. In the absence of photometric noise, these selection boxes would return relatively pure samples of mock galaxies above a given redshift limit.}
    \end{figure*}

    \begin{figure*}[ht!]
    \centering
    \includegraphics[width=\textwidth]{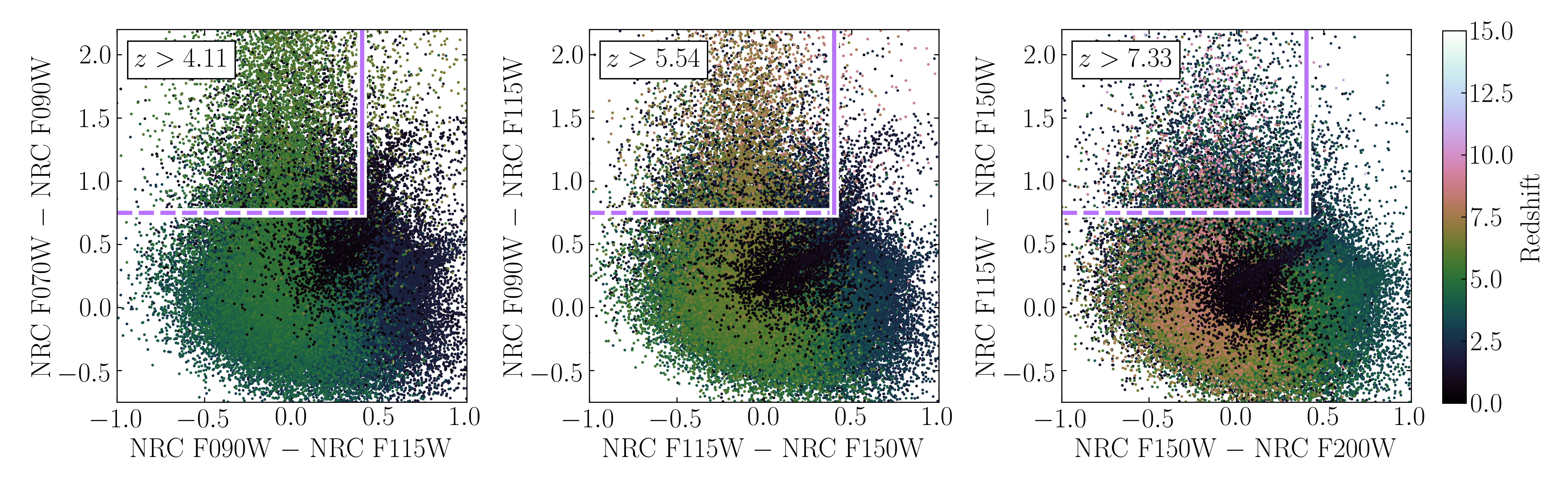}
    \caption{
    \label{fig:jaguarnoisycolor} NIRCam color-color plots for a 10'x10' section of the JAGUAR catalog, with mock galaxies at $z = 0.2 - 15$, created by simulating noise from images with a total of 98.9 ksec exposure time in each filter. In each panel, we only plot mock galaxies with detections in filters at wavelengths longer than the Lyman break with a SNR $> 3$. We show an example two-color selection method used throughout this work in lavender. In Section \ref{sec:results}, we will discuss the properties of mock galaxies selected using this selection method, where we fix the Lyman break cut (dashed line) and vary the UV continuum cut (solid line).}
    \end{figure*}

\subsection{The JAGUAR Catalog}
\label{sec:JAGUAR}

The JAGUAR mock catalog consists of a series of 11' by 11' photometric and spectroscopic catalogs, as described in \citet{williams2018}. JAGUAR includes both quiescent and star-forming mock galaxies using as the base catalog the observations of the galaxy stellar mass function from \citet{tomczak2014} at $z < 4$ and the UV luminosity function from \citet{bouwens2015} and \citet{oesch2018} at $z > 4$. These mass and luminosity functions are joined at $z = 4$ by modeling the evolution of the relationship between observed galaxy stellar mass and $M_{UV}$, the absolute magnitude of each galaxy in the ultraviolet, in agreement with measurements in the 3D-HST survey \citep{skelton2014}. JAGUAR mock galaxies were generated such that they followed the evolution of the mass and luminosity functions, and each object was then assigned a spectrum using BEAGLE, a tool designed to model and interpret galaxy SEDs \citep{chevallard2016}. This code allows for the creation of realistic mock galaxy SEDs with self-consistent nebular continuum and line emission. A large quantity of BEAGLE galaxy realizations was constructed across a wide parameter space, including fits to existing 3D-HST objects, and each galaxy in the mock catalog was matched to an individual SED from these realizations. For each object, simple S\'ersic profiles were assigned following observations of high-redshift galaxies in \citet{vanderwel2014}, which have been shown to agree with low-redshift results from Sloan Digital Sky Survey observations \citep{shen2003, guo2009}. We plot an example JAGUAR F070W dropout (at $z = 5.6$), F090W dropout (at $z = 7.5$), and F115W dropout (at $z = 9.8$) with the HST/ACS and \textit{JWST}/NIRCam filters in Figure \ref{fig:LyADropoutExample}.

The JAGUAR catalogs span a stellar mass range of $\log{(M_*/M_\sun)} = 6 - 12$ and a redshift range of $z = 0.2 - 15$. For the lowest mass mock galaxies, the catalog requires significant extrapolation of existing mass and luminosity functions. We refer the reader to \citet{williams2018} for a description of how the JAGUAR catalog agrees with current observations of the evolution of quiescent and star-forming galaxy properties, the cosmic star formation rate density, specific star-formation rate, and mass-metallicity relationship. The effects of IGM absorption in JAGUAR mock galaxies follow the prescription from \citet{inoue2014}. Dust attenuation of both the stars and the photoionized gas in the JAGUAR mock galaxies is described using a two-component model of \citet{charlotfall2000} and parameterized using $\hat{\tau}_V$, the total attenuation optical depth which is allowed to vary between 0 and 4, and the fraction of attenuation arising in the diffuse ISM $\mu$, which is fixed at 0.4. While this range is motivated from observational relations \citep{schaerer2010}, current samples of high-redshift galaxies that form the basis for these relations are likely missing a population of extremely dusty star-forming galaxies which may be observed with NIRCam \citep{casey2014,spilker2016,williams2019,wang2019}. We further discuss these sources in Section \ref{sec:dustygalaxies}. 

We plot the NIRCam color space for a 10'x10' JAGUAR realization in Figure \ref{fig:jaguarcolor}, with mock galaxy points colored by their redshift. As can be seen in each panel, at specific redshifts where the filters span the Lyman break (plotted on the y-axis), the mock galaxies are observed to have redder colors. In each color selection scenario, there are also lower redshift interlopers with red colors, a mixture of those with strong 4000\AA\ +Balmer breaks, star-forming galaxies with heavy dust obscuration, and quiescent galaxies. As would be expected with the evolution of the galaxy luminosity function to higher redshifts, the density of high-redshift dropout candidates decreases from F070W dropouts to F115W dropouts. We also overlay an example two-color dropout selection box in each panel to illustrate how objects lying inside the region at the top-left of each panel could be selected as dropout candidates. 

\subsection{Generating NIRCam Noise Estimates}
\label{sec:NIRCamNoise}

While such plots as Figure \ref{fig:jaguarcolor} can be very helpful for choosing color criteria for selecting galaxies at specific redshift ranges, these plots do not incorporate any noise, which will preferentially affect fainter (and often lower mass) galaxies, moving them both into and out of color selection regions. To simulate noise we wrote a suite of custom scripts for use with the JAGUAR catalog, NIRCPrepareMock\footnote{https://github.com/kevinhainline/NIRCPrepareMock}. These scripts generate artificial noise directly from the JAGUAR photometry, which can be used when assessing the efficacy of photometric redshift or SED fitting codes. 

We estimate noise for the mock galaxies in each filter separately, starting with the JAGUAR flux in that filter, as well as the morphology of the mock galaxy. The code selects the smallest circular aperture from a series of fixed radii (0\secpoint16, 0\secpoint24, 0\secpoint32, and 0\secpoint64) that would encompass the semi-major axis half-light radius of each mock galaxy. At this point, the script calculates the total flux of each mock galaxy through that circular aperture taking into account its S\'ersic index. Because we are not extracting flux from mock PSF-convolved NIRCam images and extracting fluxes directly, we do not correct for aperture losses. To simulate the sky background, we use estimates for the zodiacal light emission in the GOODS-S region for each filter\footnote{following the \textit{JWST} background model described here: https://jwst-docs.stsci.edu/observatory-functionality/jwst-background-model}, and add this to the flux of each mock galaxy through the aperture to produce the final flux in a given exposure. The uncertainty on the flux for an individual exposure is the Poisson noise summed in quadrature with the instrument read noise (summed over the pixels in the aperture).

When using NIRCam, individual frames will be co-added to create a final deep image from which flux will be measured. To simulate this process, the code co-adds exposures (with a base exposure time), while allowing the user to input the total number of summed frames for a final exposure. To estimate the noise, we randomly sample from a Gaussian with the RMS width set to the exposure noise to produce the frame uncertainty, and then sum the noise in quadrature for each co-added frame.

There are some caveats to this approach to estimating uncertainty. Because we are not using mock images, we do not account for the instrumental point spread function, or change the size of the aperture in different filters to account for the change in instrument resolution as a function of wavelength. Accounting for the PSF would serve to decrease the flux that falls inside a given aperture at longer wavelengths, as the 50\% encircled energy (defined as the fraction of light contained inside a circular aperture) increases from 0\secpoint068 for the F070W filter to 0\secpoint092 for the F444W filter. This effect would serve to make objects artificially more blue when using longer wavelength filters, although PSF-matching can account for this effect. For the majority of the analysis presented here, we focus on the NIRCam short-wavelength filters (F070W, F090W, F115W, F150W, and F200W), where this effect is minimized. In addition, we don't simulate projection effects which would serve to reduce the number of high-redshift galaxies which are blocked by foreground galaxies. The full treatment of estimating noises from mock images is beyond the scope of the current work. While there are more rigorous methods of measuring fluxes, the flux uncertainties produced by our code agree with the predicted uncertainties from the STScI \textit{JWST} Exposure Time Calculator\footnote{http://jwst.etc.stsci.edu/}, and our code can be run quickly on large samples. 

In Figure \ref{fig:jaguarnoisycolor}, we plot the same NIRCam color spaces as in Figure \ref{fig:jaguarcolor}, but with fluxes from a 100 square arcminute noisy catalog with images at 49.5 ksec total exposure time. We only plot mock galaxies detected with SNR $> 3$ in the filters on the x-axis in each panel (we do not set a limit on the SNR for the dropout filter). By comparing the noise-free to the noisy photometry, we can observe how rare dropout candidates are at $z > 8$, even in a 100 square arcminute field, both because of the faint observed fluxes of these objects (less mock galaxies satisfy the SNR $> 3$ criterion), as well as their low on-sky density based on the observed UV luminosity functions used to constrain JAGUAR. We also plot the same selection boxes as in Figure \ref{fig:jaguarcolor}, demonstrating the difficulty in separating high redshift targets and lower redshift interlopers with noisy photometry.

\subsection{Mock Survey Design}
\label{sec:SurveyDesign}

\begin{figure}[t!]
    \centering
    \includegraphics[width=\columnwidth]{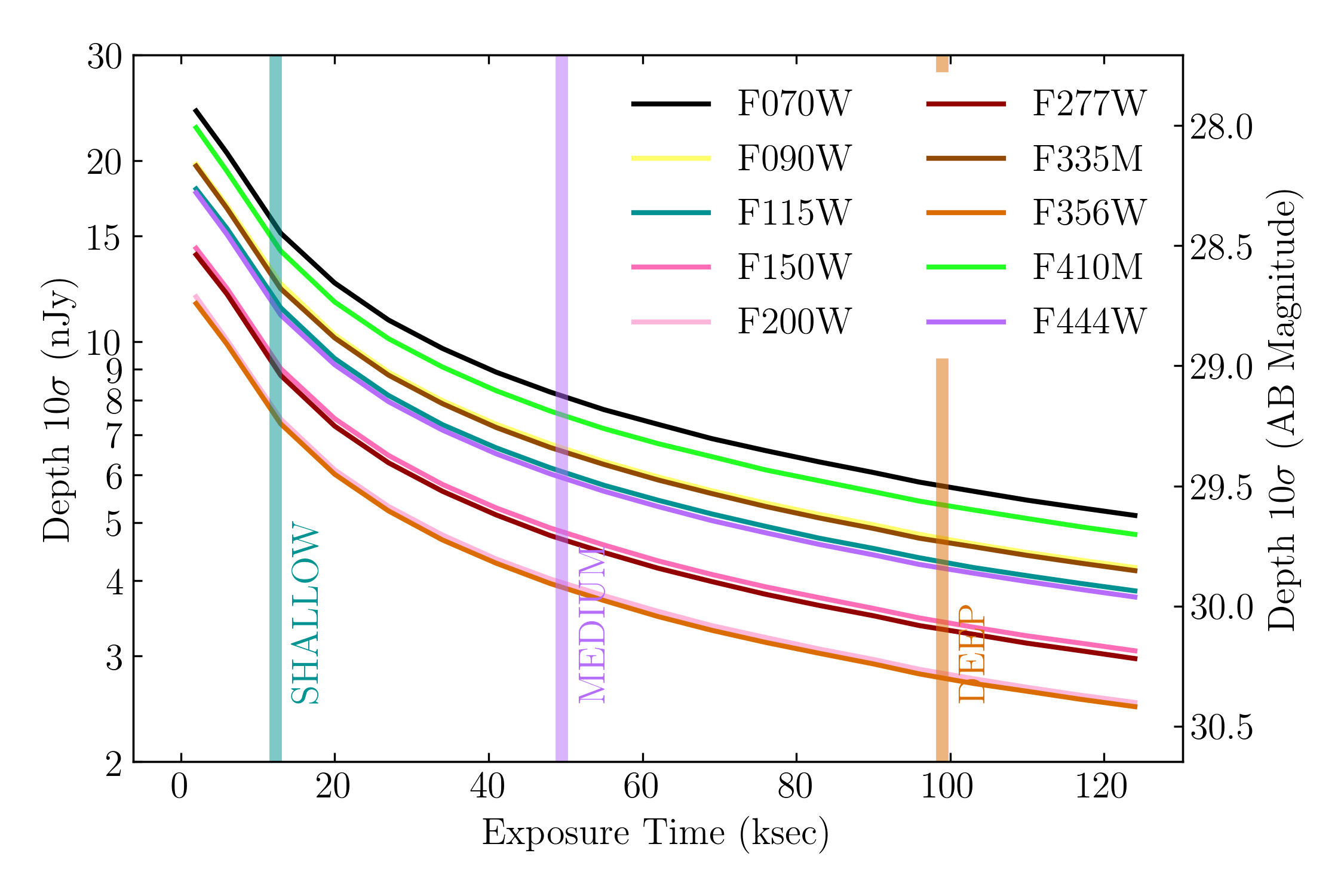}
	\caption{Simulated 10$\sigma$ depths plotted against total exposure time for the NIRCam filters used in this work. These values were estimated using the NIRCPrepareMock package. We also plot the exposure times for the SHALLOW (teal), MEDIUM (lavender), and DEEP (orange) surveys with vertical lines.
	\label{fig:DepthVsExposureTime} } 
    \end{figure}

To explore high-redshift dropout selection with NIRCam and \textit{HST}+NIRCam, we generated multiple sets of mock catalogs with realistic noise estimates. Because NIRCam may target regions of the sky that do not have adequate deep \textit{HST} coverage, we produced noisy data sets with only NIRCam coverage over a region of 100 square arcminutes, with three different depths. In each case, we simulated a  \textit{JWST}/NIRCam observational strategy for observing high-redshift galaxies which utilizes the DEEP8 readout pattern, with 7 groups per integration, for a pixel integration time of 1374.3 seconds. For each depth we assumed a 9-point dither pattern, which samples 3 times the pixel resolution, and we then varied the number of integrations per exposure: \begin{enumerate}
    \item A "SHALLOW" mock survey with 1 integration per exposure resulting in an integrated exposure time of 12.3 ksec per filter.
    \item A "MEDIUM" mock survey with 4 integration per exposure resulting in an integrated exposure time of 49.5 ksec per filter.
    \item A "DEEP" mock survey with 8 integration per exposure resulting in an integrated exposure time of 98.9 ksec per filter.
\end{enumerate} 
We plot the median 10$\sigma$ depths in each of the NIRCam filters we will use in this study as a function of total exposure time in Figure \ref{fig:DepthVsExposureTime}. These depths were calculated from our simulated noisy photometry and are appropriate for extended sources. Future \textit{JWST}/NIRCam deep surveys will likely be designed with longer exposures in less sensitive bands in order to balance the observational depth, and interested readers can explore the impact of such changes with the NIRCPrepareMock code we make publicly available. Deeper observations at bluer NIRCam bands will preferentially affect the ability for a given survey to remove low redshift interlopers, while deeper observations in the detection bands for a given selection criterion will lead to a larger number of recovered high-redshift objects. More exposure time in longer-wavelength NIRCam bands will be important for SED fitting, as these bands cover the rest-frame optical and a suite of strong emission lines in high-redshift galaxies.

\begin{deluxetable}{lrrr}
\tablecaption{Simulated NIRCam 10$\sigma$ depths for the SHALLOW, MEDIUM, and DEEP surveys. \label{tab:SurveyDepths}}
\tablehead{
 & \multicolumn3c{10$\sigma$ Depth (nJy)}  \\
\colhead{Filter} & \colhead{SHALLOW} & \colhead{MEDIUM} & \colhead{DEEP} 
}
\startdata
F070W & 15.71 & 8.13 & 5.76 \\
F090W & 12.93 & 6.67 & 4.71 \\
F115W & 11.79 & 6.08 & 4.30 \\ 
F150W & 9.35 & 4.83 & 3.42 \\ 
F200W & 7.69 & 3.98 & 2.81 \\
F277W & 9.11 & 4.70 & 3.32 \\
F335M  & 12.69 & 6.58 & 4.65 \\
F356W & 7.56 & 3.91 & 2.76 \\
F410M & 14.67 & 7.56 & 5.76 \\
F444W & 11.48 & 5.95 & 4.20
\enddata
\end{deluxetable}

In addition, in order to explore how NIRCam observations can complement those made at shorter wavelengths by \textit{HST}, we simulate a region of the sky of 10.8 square arcminutes at the XDF ACS depth given by \citet{illingworth2013}. We simulate observations in the \textit{HST}/ACS filters F435W (152.4 ksec, 7.06 nJy 10$\sigma$ depth), F606W (174.4 ksec, 5.00 nJy 10$\sigma$ depth), F775W (377.8 ksec, 5.99 nJy 10$\sigma$ depth), F814W (50.8 ksec, 21.93 nJy 10$\sigma$ depth), and F850LP (421.6 ksec, 10.61 nJy 10$\sigma$ depth), and generate NIRCam fluxes with the same depths as described in the previous paragraph, but over the smaller XDF area. For both mock surveys, we produced 500 noisy samples to explore how our noise estimates affect the uncertainties on the overall density of objects selected by a set of NIRCam color cuts. 

\subsection{Color Cut Figures of Merit}
\label{sec:ColorCutStatistics}

Because of the large variety of observed galaxy SEDs and photometric noise, there is no single ideal set of color selection criteria that will result in a clean sample of high-redshift galaxies. Our goal in this paper is to estimate statistics on the recovered population of simulated high-redshift galaxies as a function of our color cuts in order to aid in future NIRCam observations. For the purposes of this study, we require a definition of a "high-redshift object" and an "interloper" for a given dropout selection filter. While the Lyman limit is found at 912\AA, absorption due to the Ly$\alpha$ forest causes the exact wavelength of the Lyman break to shift to longer wavelengths at higher redshifts, which is simulated within the JAGUAR catalog. At $z > 6$, this absorption is thick enough that the break occurs at 1216\AA, the wavelength of Ly$\alpha$. We define a high-redshift object as one that is above the redshift where the Ly$\alpha$ emission line crosses the half-power response of the blue side of the dropout band, and an interloper is any object that satisfies a given color selection criteria, but is below this redshift. 

There are three primary statistics that we explore for choosing a given color selection criterion and assembling a high-redshift dropout sample:

\begin{enumerate}
    \item The first statistic we report is selection ``accuracy,'' defined as the ratio between the number of high-redshift objects selected to the \textit{total} number of objects selected by a given color selection criterion. This is sometimes referred to as sample ``purity'' in the literature. 
    \item Extremely red selection limits will result in more accurate, but smaller total samples, so we also report the on-sky density of high-redshift objects under a given selection criterion, which we refer to as ``true positive density'', or TPD. 
    \item The final statistic we provide is selection ``completeness,'' defined as the ratio between the number of high-redshift objects selected to the number of high-redshift galaxies that satisfy the SNR criteria (both red detections and blue non-detections). 
\end{enumerate}

Defining the optimal selection criteria will be determined by the trade-off between a more accurate sample, and one that has a higher number of high-redshift objects selected and a higher sample completeness. 

    \begin{figure*}[ht!]
    \centering
    \includegraphics[width=0.85\textwidth]{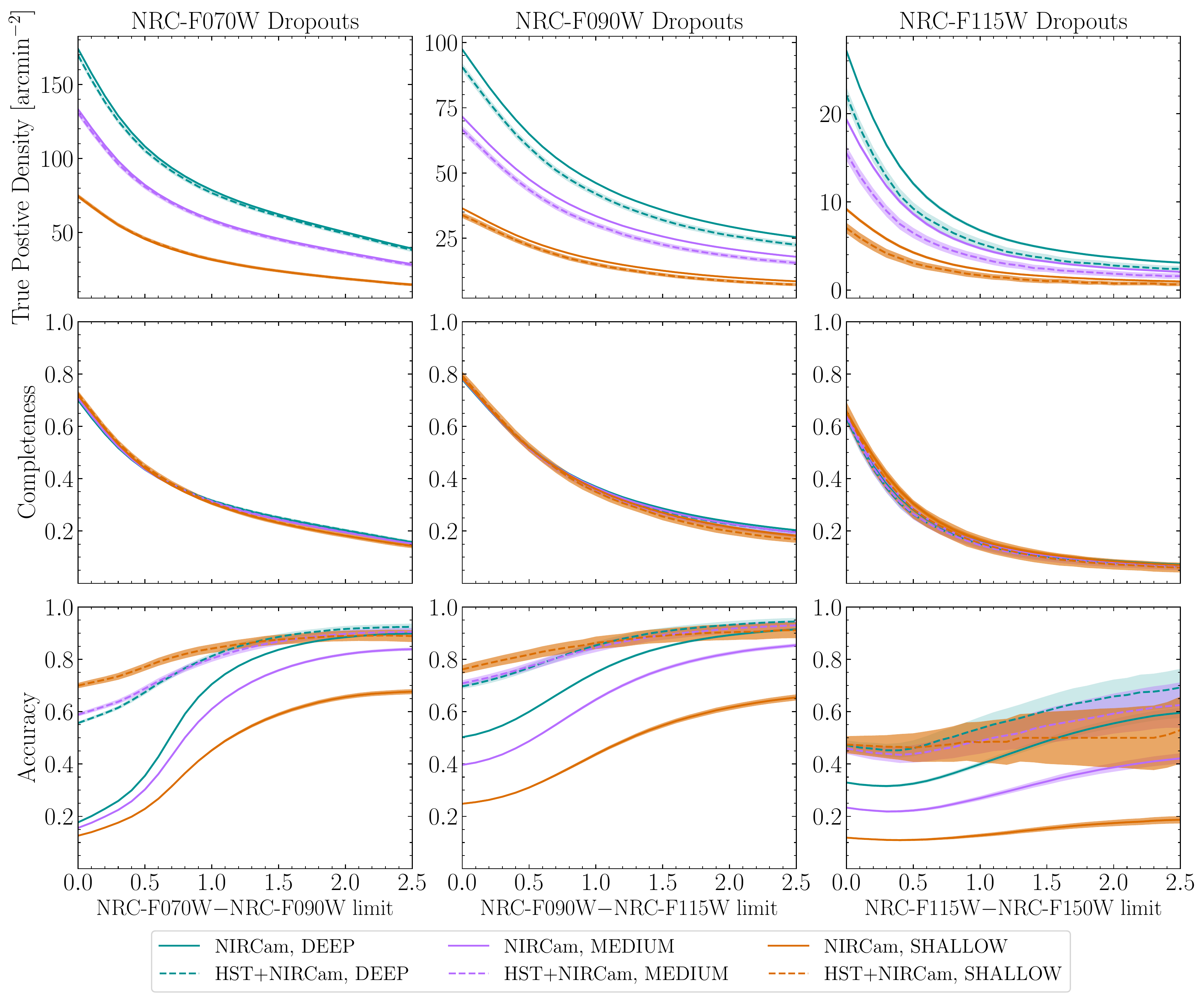}
	\caption{TPD (top row), completeness (middle row), and accuracy (bottom row) as a function of color cut for F070W dropouts (left), F090W dropouts (middle), and F115W dropouts (right), for the DEEP (teal), MEDIUM (lavender), and SHALLOW (orange) NIRCam (solid) and \textit{HST}+NIRCam (dashed) surveys requiring a SNR $> 3$ in both detection bands. We additionally require a second color cut of F090W - F115W $< 0.4$ (left), F115W - F150W $< 0.4$ (middle), and F150W - F200W $< 0.4$ (right). The shaded regions indicate the 1$\sigma$ range on the TPD and accuracy values calculated using the 500 noisy mock catalogs. In each set of panels, using a redder color cut results in lower TPD and completeness at a higher level of accuracy, and requiring a higher SNR limit reduces the overall TPD while increasing the accuracy at a given color cut. 
	\label{fig:DepthComparison}} 
    \end{figure*}

\subsection{Selecting High-Redshift Galaxies}
\label{sec:selectinggalaxies}

High-redshift dropout candidates are often selected by observing flux at a given significance in multiple photometric filters at wavelengths longer than the break, with flux below a given significance at wavelengths shorter than the break. In this paper, we select mock galaxies based on a set of color criteria, and require objects to be selected in at least two filters to the red of the Lyman break above a signal-to-noise ratio (SNR) of 3 (although we will describe how our statistics change if we instead select above a SNR of 5, or 10). In addition, because of IGM absorption at rest wavelengths shorter than the Lyman break, we require a non-detection in the bands to the blue of the dropout filter at a SNR less than 2, as is commonly used in the literature \citep[e.g.,][]{bouwens2015}. We should note that for the NIRCam-only simulations we will describe, F070W dropouts will not have a rejection band, while for F090W dropouts we will use F070W to help reject interlopers, and for F115W dropouts, we require non-detection fluxes in both F070W and F090W. For the \textit{HST}+NIRCam simulations we will also use the \textit{HST} bands for this rejection, highlighting the importance of using shorter wavelength data for selecting more pure samples of objects with fewer lower redshift interlopers. For F070W dropouts, we require non-detections at \textit{HST}/ACS F435W. For F090W dropouts, we require non-detections at \textit{HST}/ACS F435W, F606W, and NIRCam F070W. For F115W dropouts, we require non-detections at \textit{HST}/ACS F435W, F606W, F775W, F814W, and NIRCam F070W and F090W. 

\section{Results}
\label{sec:results}

In this section we discuss our number density, completeness, and accuracy results as a function of multiple factors, including survey depth, detection SNR, and survey design. For the majority of this analysis, we will adopt a simple two-color cut selection method, as illustrated in Figures \ref{fig:jaguarcolor} and \ref{fig:jaguarnoisycolor}, where we vary the color cut for the filters that straddle the Lyman break (the ``Lyman break cut'', represented by a dashed line in these figures), and we fix the color requirement for the filters redward of the Lyman break (the ``UV continuum cut'' represented by a solid line in these figures). After testing the effects of varying the UV continuum cut on TPD and accuracy, we require F090W - F115W < 0.4 (magnitudes, for F070W dropouts), F115W - F150W < 0.4 (for F090W dropouts), and F150W - F200W < 0.4 (for F150W dropouts). We will be discussing the use of single color cuts or more complicated color selection methods further in Section \ref{sec:xcuts}. 

\subsection{Survey Depth and Detection SNR}
\label{sec:surveydepth}

The design of a survey, and especially the observational depth in the chosen filters, will have a strong impact on the number of high-redshift objects that are recovered with a given selection method. In Figure \ref{fig:DepthComparison}, we plot TPD (top panels), completeness (middle panels), and accuracy (bottom panels) against the Lyman break cut for our DEEP, MEDIUM, and SHALLOW survey depths. In each set of panels we utilize a detection SNR of 3.0 (for at least two filters to the red of the Lyman break), and ensure non-detections in the filters to the blue of the Lyman break as previously described. We plot the NIRCam only selection with solid lines, and the \textit{HST}+NIRCam selection with dashed lines. We plot the 1$\sigma$ range on the distribution of these values calculated using the 500 mocks with a shaded region.

    \begin{figure*}[ht!]
    \centering
    \includegraphics[width=0.95\textwidth]{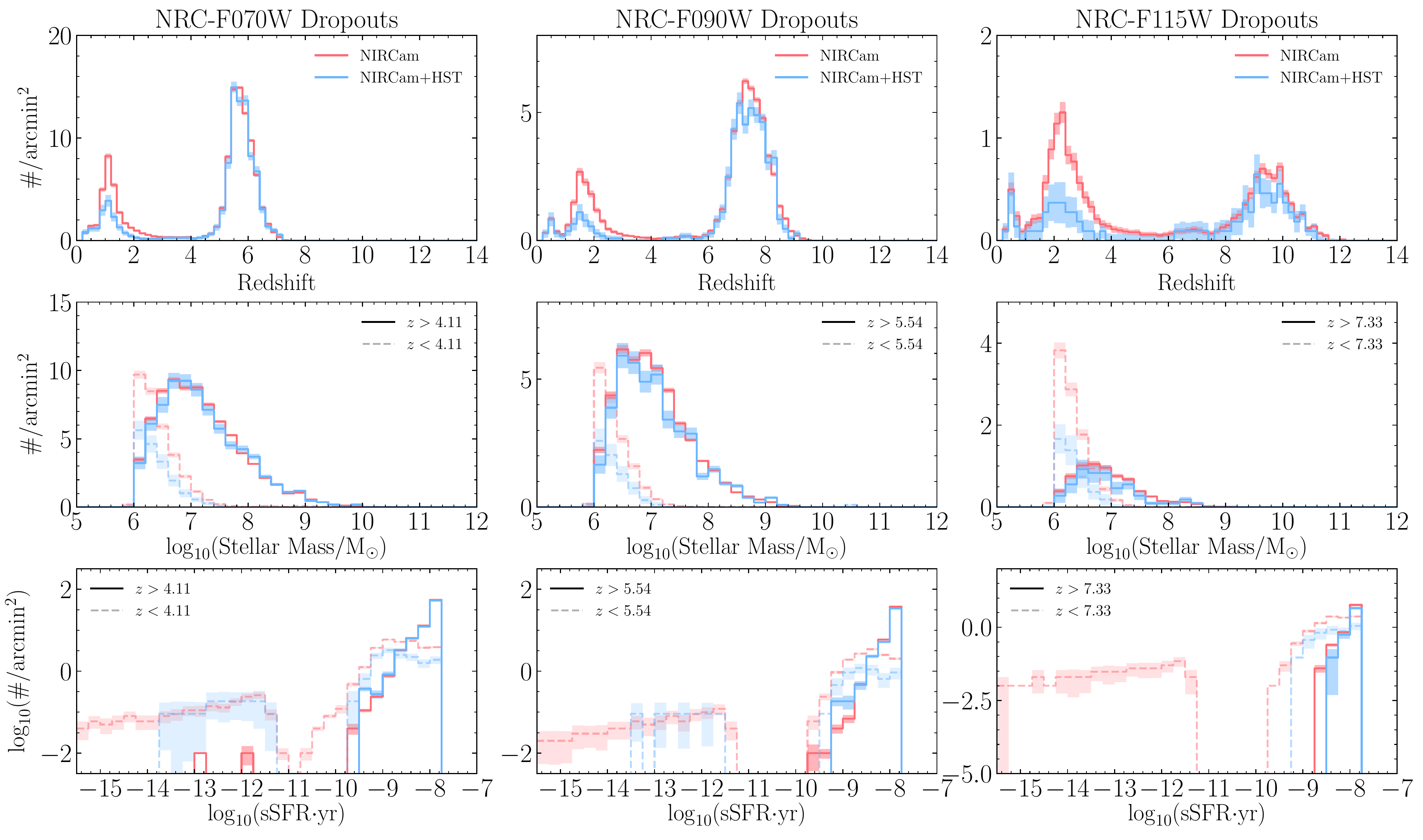}
	\caption{
	\label{fig:specmasshistograms} Histograms of mock galaxies at the DEEP survey depth with F070W - F090W $> 1.0$ (left), F090W - F115W $> 1.0$ (middle), and F115W - F150W $> 1.0$ (right), as well as a second color cut as described in Figure \ref{fig:DepthComparison}, with a detection SNR > 3 for all three plots. (Top) The spectroscopic redshifts of the dropouts, where in red we plot the number density of objects with NIRCam data alone, and in blue we plot those objects with \textit{HST}+NIRCam data. The addition of deep \textit{HST} data for constraining blue non-detections has a significant effect in removing interlopers. (Middle) The JAGUAR stellar masses of these objects are plotted, with red and blue as in the top panel, but now the dashed lines correspond to a subsample of low-redshift interlopers in each panel, while we plot the mass distribution of the true high-redshift objects with a solid line. (Bottom) The JAGUAR specific star formation rates (sSFR) are plotted similar to the middle row, but using a logarithmic scale on both axes. Interlopers are primarily found at low stellar masses, although for the F070W and F090W dropouts, a number of higher-mass quiescent mock galaxies are selected as interlopers with these color cuts, which are are also found at lower sSFR values.}
    \end{figure*}

For all three color selection criteria, at redder color cuts the density and completeness of recovered true high-redshift sources decreases, but the accuracy of the sample increases. The total number of recovered sources, as well as the overall accuracy, increases at deeper survey depths. The completeness, however, does not depend strongly on survey depth, as this statistic is a ratio between two values that depend on depth in roughly the same manner. At a detection SNR > 3, it is only possible to reach high levels of accuracy with extremely red color cuts. In all survey depths and dropout criteria, the accuracy plateaus to a value less than 1.0 owing to contamination by mock galaxies at low redshifts and low SNR with non-detections in the bluer filter of the Lyman break color cut. As a result, these objects have extremely red Lyman break colors, and would be contaminants at any choice of cut. We should also note that the 1$\sigma$ distributions are much larger for the NRC-F115W dropouts in the SHALLOW depth survey because of the small number of objects recovered at this survey depth.

While the usage of \textit{HST} blue filter non-detections results in overall lower densities of actual high-redshift objects, as would be expected, it has a much larger effect on the accuracy. For F090W - F115W $> 1.0$, in the DEEP survey, with NIRCam observations only, the density of sources at $z > 5.5$ is 50 arcmin$^{-2}$, at an accuracy of 0.70, while with \textit{HST}+NIRCam observations, the density is 10\% smaller, but at an increased accuracy of 0.80. Interestingly, when using HST fluxes, the measured accuracy at blue color cuts for the SHALLOW depth survey is higher than for the MEDIUM or DEEP surveys. The addition of an HST SNR cut has a strong effect on reducing the total number of galaxies selected by a set of color cuts (and thereby increasing accuracy) which is more significant in the SHALLOW survey due to the larger flux uncertainties. The discrepancy between the NIRCam and the \textit{HST}+NIRCam TPD values is larger for higher-redshift dropouts, because of the additional blue filters that are used to reject low-redshift interlopers. The recovered completeness is not significantly different between \textit{HST}+NIRCam and NIRCam observations only. 

    \begin{figure*}[ht!]
    \begin{center}
    \includegraphics[width=0.85\textwidth]{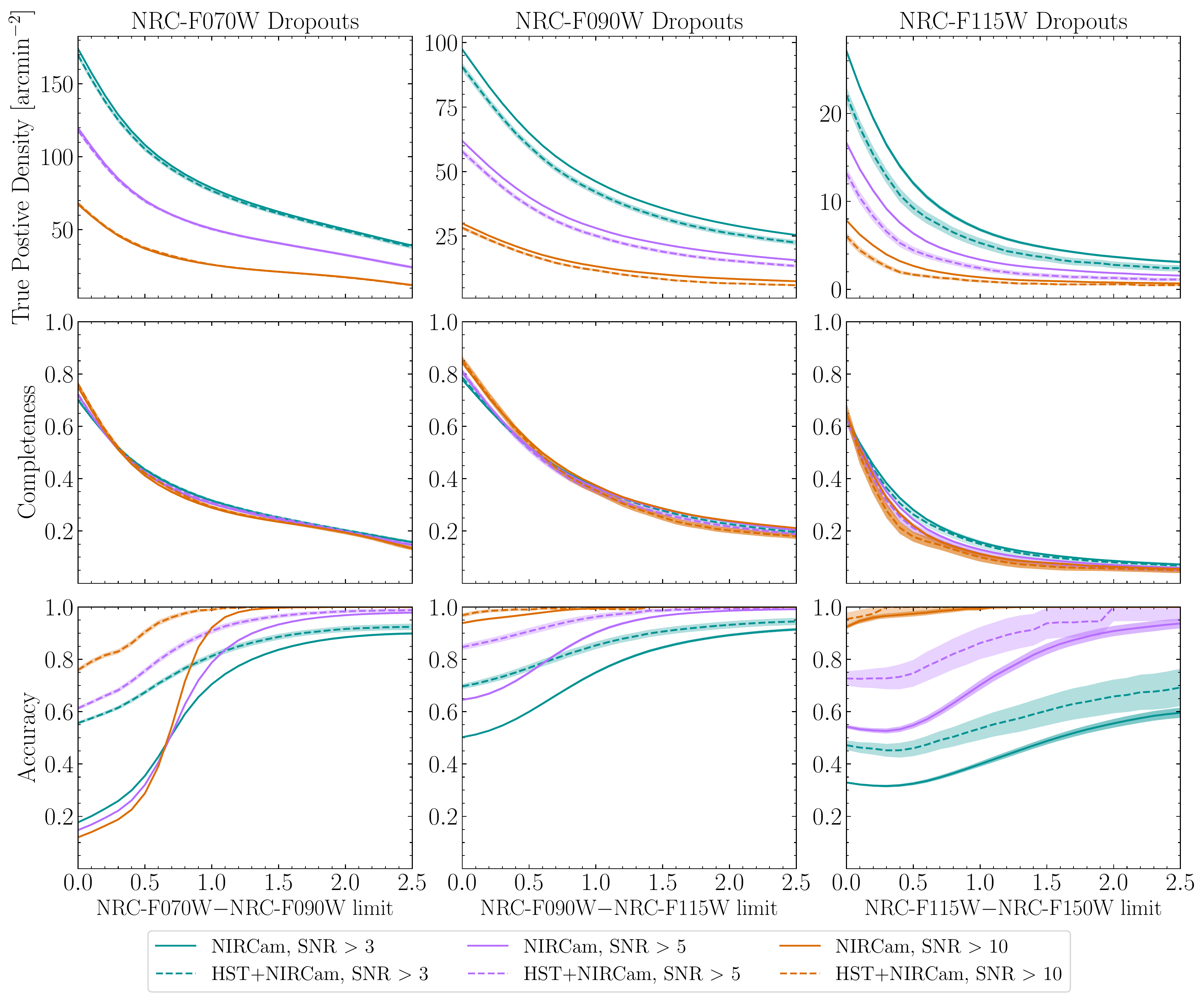}
	\caption{\label{fig:SNRComparison} TPD (top row), completeness (middle row), and accuracy (bottom row) as a function of color cut for F070W dropouts (left), F090W dropouts (middle), and F115W dropouts (right), for a detection SNR of $>3$ (teal), $>5$ (lavender), and $>10$ (orange) for NIRCam (solid) and \textit{HST}+NIRCam (dashed) DEEP surveys. We additionally require a second color cut of F090W - F115W $< 0.4$ (left), F115W - F150W $< 0.4$ (middle), and F150W - F200W $< 0.4$ (right). In each set of panels, a higher SNR restriction leads to an increase in the accuracy, but at a significant decrease in TPD.} 
    \end{center}
    \end{figure*}

We can examine in more detail the properties of the mock galaxies that are recovered by a specific color cut. In Figure \ref{fig:specmasshistograms}, we show the redshift (top), stellar mass (middle), and specific star formation rate (sSFR, defined as the mock galaxy star formation rate normalized by the stellar mass, bottom) distributions for the mock galaxies selected by color cuts of F070W - F090W $> 1.0$ (left), F090W - F115W $> 1.0$ (middle), and F115W - F150W $> 1.0$ (right) (in addition to the UV continuum cuts described above) for the DEEP survey. In all three columns, we plot the NIRCam-only selection in red, and the \textit{HST}+NIRCam selection in blue. In each case, we can see how mock galaxies at $z \sim 1 - 4$ are the primary contaminants, and based on the mass distributions, these objects have masses 10$^6$ - 10$^7$ M$_{\sun}$ and lower sSFR values. The addition of the \textit{HST} data helps mitigate the contaminants, but in all cases, red, low-mass, faint mock galaxies are selected as Lyman-break galaxies.  

We also explored how the detection SNR affects the dropout selection. For the DEEP survey depth, using the NIRCam and \textit{HST}+NIRCam observations, we calculated the TPD, completeness, and accuracy for detection SNR of 3, 5, and 10, and plot these results in Figure \ref{fig:SNRComparison} for the F070W, F090W, and F115W dropouts. Changing the SNR has a strong effect on the accuracy of the recovered samples, such that samples with greater than 90\% accuracy can be recovered with a detection SNR of 5 - 10 at redder color cuts. However, this comes at a significant cost to the recovered TPD: almost twice as many objects are detected for a detection SNR of 3 vs.\ 5, and at 5 vs.\ 10 at all color limits. While the completeness is similar between the different SNR cuts for F070W and F090W dropouts, we find a slightly higher completeness for the detection SNR of 3 for F115W dropouts.  

    \begin{figure}[ht!]
    \centering
    \includegraphics[width=\columnwidth]{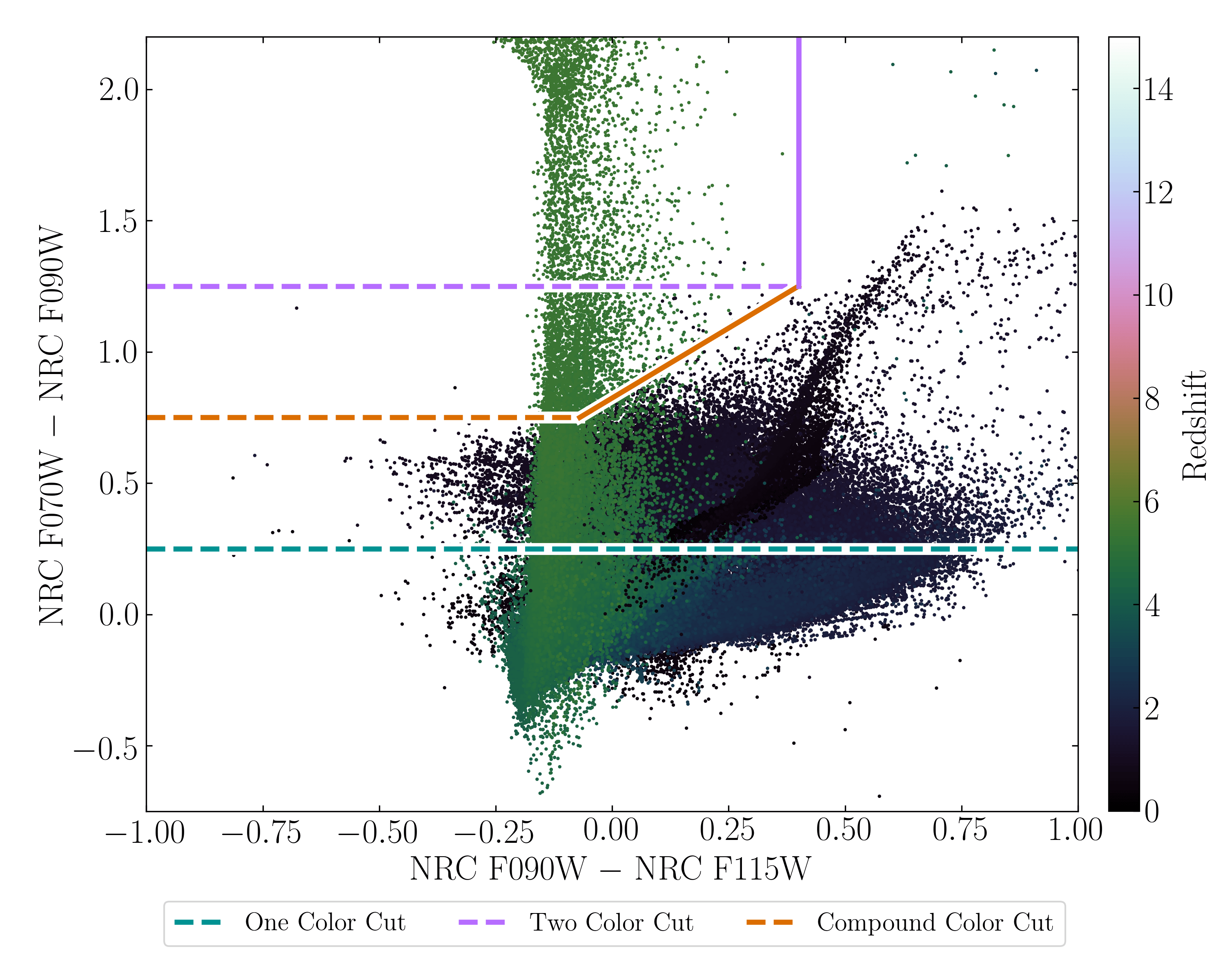}
	\caption{
	\label{fig:colorcutcomparison} NIRCam color-color plot with the three selection criteria that we employ and compare in this paper. In lavender, we show the two-color selection method used in Section \ref{sec:surveydepth}. In teal and orange, we show the One Color and Compound Color Cut selection methods, respectively.}
    \end{figure}

\subsection{Single Color Cut vs. Two Color Cut Selection}
\label{sec:xcuts}

Throughout this analysis, we have shown results with a simple two-color selection method, as star-forming mock galaxies at lower redshift can have red colors which mimic dropout galaxies. In this section we additionally explore the recovery of high-redshift galaxies with a method which uses only a single color cut, as well as a compound method which uses two color cuts and an additional angled color selection as is often used to select high redshift galaxies. In Figure \ref{fig:colorcutcomparison}, we show the F070W dropout selection color space marked to show the two-color selection we have used up to this point (lavender), single-color selection (teal), and the compound color selection (orange). Because dust obscuration in a galaxy results in redder colors, for each dropout selection criteria, the slope of our angled cut corresponds to the reddening vector for the \citet{calzetti2000} dust prescription for that filter combination\footnote{For F070W dropouts we use a slope of 1.07 and an intercept of 0.82, for F090W dropouts we use a slope of 1.03 and an intercept of 0.84, and for F115W dropouts we use a slope of 1.07 and an intercept of 0.82.}. To simulate the different selection methods, we repeated our previous analysis using these alternate selection methods on the DEEP survey, with a detection SNR of 3.0, but we fix the solid lines shown in Figure \ref{fig:colorcutcomparison} and explore how changing the color indicated by the dashed lines impacts the recovery of high-redshift galaxies.

    \begin{figure*}[ht!]
    \centering
    \includegraphics[width=0.85\textwidth]{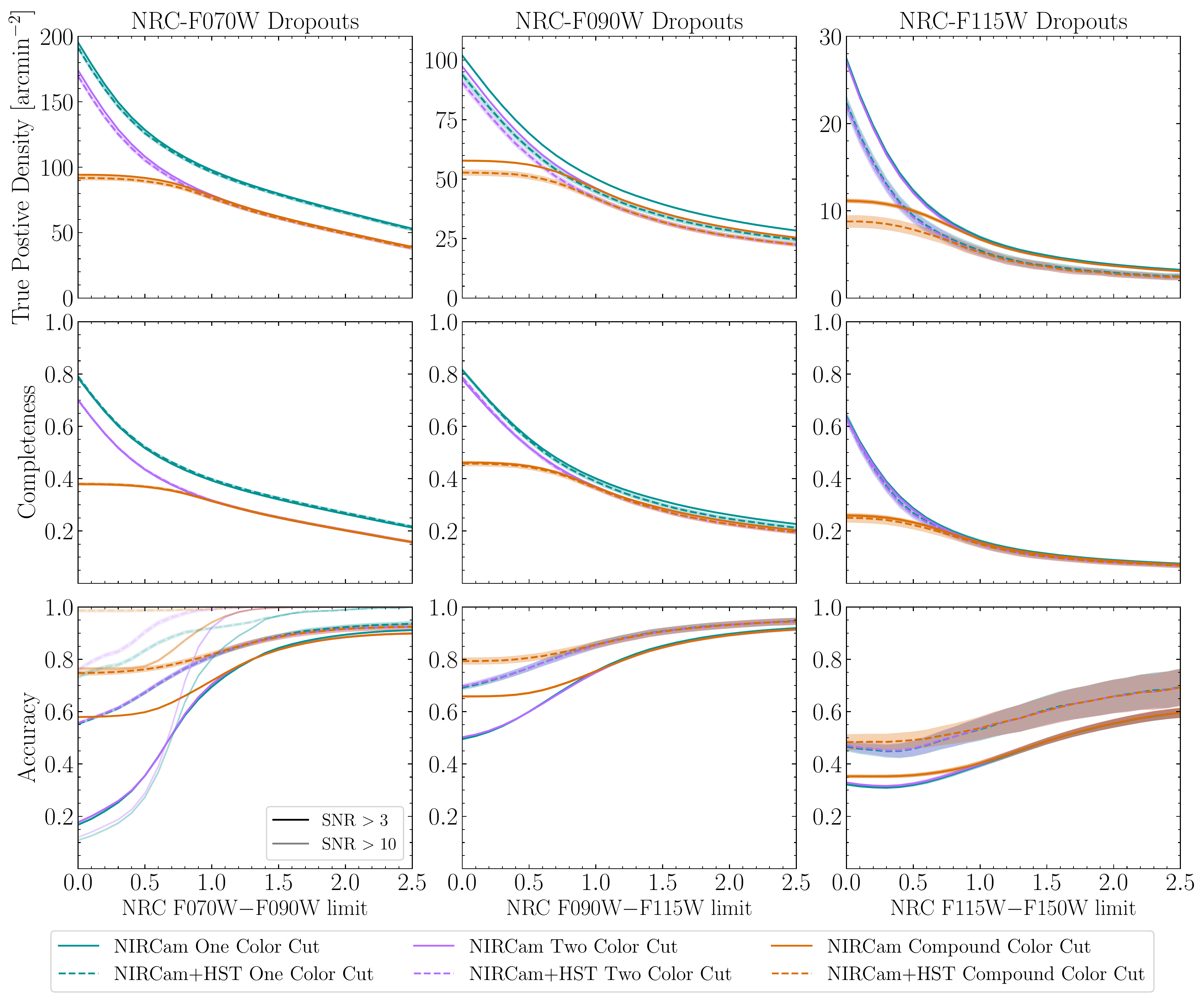}
    \caption{\label{fig:colorcutmethod} TPD (top row), completeness (middle row), and accuracy (bottom row) as a function of color cut for F070W dropouts (left), F090W dropouts (middle), and F115W dropouts (right) with the One Color Cut (teal), Two Color Cut (lavender), and Compound Color Cut (orange) methods for NIRCam (solid) and \textit{HST}+NIRCam (dashed) surveys. Above a given Lyman break color cut, the Two Color and Compound Color cut results become identical. At bluer cuts, the Compound Color selection has a higher accuracy but a lower TPD and completeness. In general, the One Color selection has a lower accuracy, except at the reddest cuts. This is clearly seen at higher detection SNR, as illustrated by comparing the accuracy at SNR > 3 and SNR > 10 for F070W dropouts in the bottom-left panel.} 
    \end{figure*}

In Figure \ref{fig:colorcutmethod}, we plot the TPD, completeness, and accuracy for F070W, F090W, and F115W dropouts, comparing the three color cut methods as shown in Figure \ref{fig:colorcutcomparison}. Not surprisingly, the One Color Cut method leads to a larger TPD for all three dropout selection techniques, as fewer objects are excluded. The completeness for the three selection methods is very similar to the TPD, in that the One Color Cut method results in larger completeness at all color cuts. The accuracy values of the One Color Cut and Two Color Cut methods are very similar for all three dropout selection techniques, likely due to the SNR $> 3$ detection threshold. At such a low SNR value, the large noise scatter on the mock galaxy colors leads to similar accuracy levels with or without the UV continuum cut. If we use a detection SNR $> 10$, the accuracy for the Two Color Cut method is larger at all color limits than that for the One Color Cut Method, as shown for F070W dropouts in the bottom-left panel of Figure \ref{fig:colorcutmethod}. The Compound Color Cut method results in the highest accuracy levels at bluer color cuts, but the third angled cut removes a significant fraction of high-redshift mock galaxies for all three dropout methods. We also find that the using a Compound Color Cut does remove relatively brighter ($m_{\mathrm{AB,F115W}} < 27$) interloper galaxies but in addition a number of faint ($m_{\mathrm{AB,F115W}} \sim 30$) high-redshift galaxies also are culled. The impact of the angled color cut on accuracy is lessened at higher redshift, where there are less dusty and quiescent mock galaxies in the JAGUAR catalog, and less reason for using a third color cut. The key result from this analysis is that for F070W and F090W dropouts, it is possible to get a significant number of candidates with an accuracy level greater than 70\% by employing a Compound Color Cut. 
        
\begin{figure*}[ht!]
    \centering
    \includegraphics[width=0.85\textwidth]{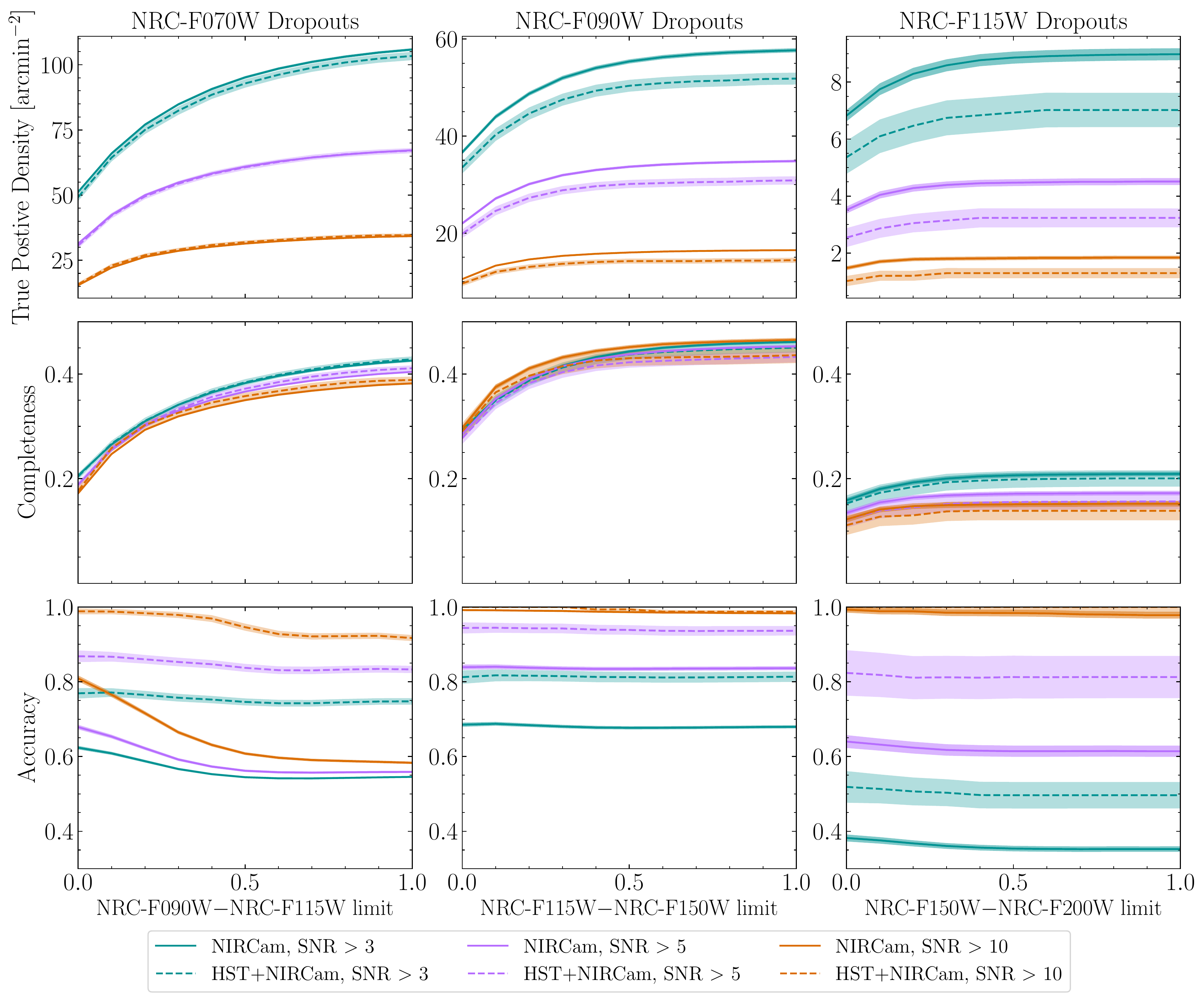}
	\caption{\label{fig:SNRxtest} TPD (top row), completeness (middle row), and accuracy (bottom row) as a function of second color cut for F070W dropouts (left), F090W dropouts (middle), and F115W dropouts (right) as a function of detection SNR; SNR > 3 (teal), SNR > 5 (lavender), and SNR > 10 (orange), for NIRCam (solid) and \textit{HST}+NIRCam (dashed) surveys. In each panel, we fix the first color cut to F070W - F090W > 1.0 (left panel), F090W - F115W > 1.0 (middle panel), and F115W - F150W > 1.0 (right panel).} 
    \end{figure*}

We have shown results using the Two Color Cut method with a fixed UV continuum cut of $<0.4$. To explore how changing this second color cut affects the resulting TPD, completeness, and accuracy, we looked at selecting high-redshift dropout candidates by fixing the Lyman break cut and varying the UV continuum color cut (In Figure \ref{fig:colorcutcomparison}, this would amount to fixing the dashed lavender line and changing the solid lavender line). For this analysis, we set F070W - F090W > 1.0, F090W - F115W > 1.0, and F115W - F150W > 1.0, and looked at mock galaxies at the DEEP survey depth. We show how TPD, completeness, and accuracy vary with the second color limit and the detection SNR in Figure \ref{fig:SNRxtest}. 

In these plots, we show that while TPD and completeness increases as the color cut becomes more inclusive, the accuracy falls, especially for F070W dropouts, due to the larger number of low-redshift interlopers. Because of these results, we have adopted a uniform color cut in our Two Color Method selection of F090W - F115W < 0.4 for F070W dropouts, F115W - F150W < 0.4 for F090W dropouts, and F150W - F200W < 0.4 for F115W dropouts. These color cuts correspond to UV slope $\beta < -0.52$ for F070W dropouts, $\beta < -0.61$ for F090W dropouts, and $\beta < -0.70$ for F115W dropouts.

\subsection{The Impact of Ly$\alpha$ Emission on Color Selection}
\label{sec:LyAEmission}

The presence of the Ly$\alpha$ emission line can contribute flux to the filters used in selecting high-redshift galaxies, potentially impacting the numbers of galaxies that are recovered by a given cut. In the JAGUAR mock catalog, the median Ly$\alpha$ rest-frame Equivalent Width (EW) is 74 \AA\ for mock galaxies at $z > 4.1$, the redshift where Ly$\alpha$ enters the NIRCam F070W filter. At $z = 7$, an emission line with this EW would result in F090W magnitude difference of $\Delta m_{AB} = 0.29$. 

\begin{figure*}[ht!]
    \centering
    \includegraphics[width=0.85\textwidth]{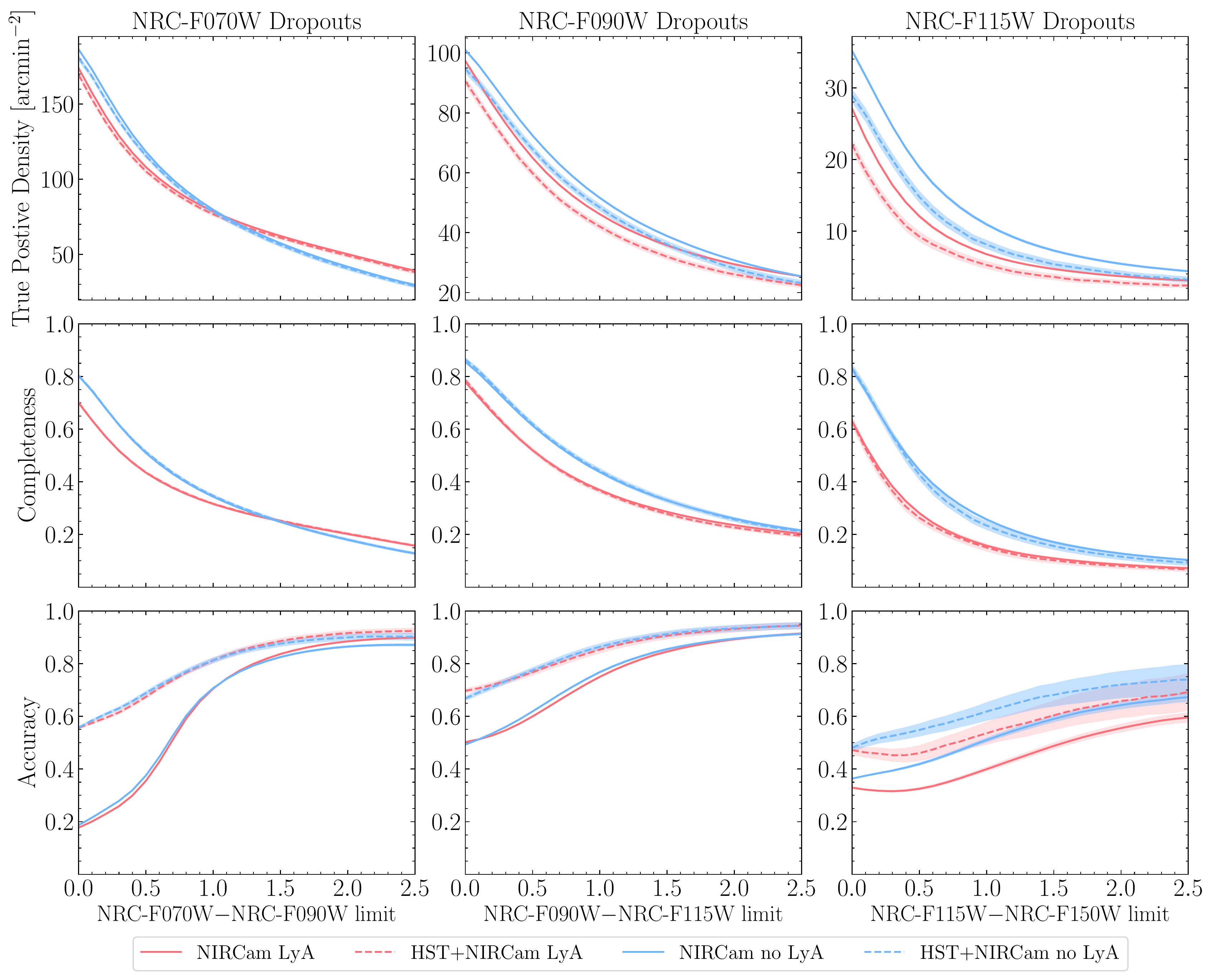}
	\caption{\label{fig:LyAComparison} TPD (top row), completeness (middle row), and accuracy (bottom row) as a function of color cut for F070W dropouts (left), F090W dropouts (middle), and F115W dropouts (right), with (red) and without (blue) Ly$\alpha$ emission for NIRCam (solid) and \textit{HST}+NIRCam (dashed) surveys. We additionally require UV continuum color cut of F090W - F115W $< 0.4$ (left), F115W - F150W $< 0.4$ (middle), and F150W - F200W $< 0.4$ (right). Ly$\alpha$ emission results in selection with a lower TPD and completeness at redder color cuts, and a higher TPD and completeness at bluer color cuts. The accuracy is similar between the two catalogs, except for F115W dropouts.} 
    \end{figure*}

Ly$\alpha$ is a resonant line, and its emission is highly dependent on the geometry of the gas in the galaxy as well as the surrounding IGM \citep{neufeld1991, giavalisco1996, kunth1998, frye2002, shapley2003}, although this resonance is not modeled for the JAGUAR mock catalog galaxies. At $z > 6$, the IGM has been observed to be increasingly neutral which significantly reduces the fraction of galaxies with observed Ly$\alpha$ in emission \citep{stark2010, pentericci2011, pentericci2014, caruana2012, caruana2014, schenker2012, schenker2014, treu2013, tilvi2014}. To explore how Ly$\alpha$ emission affects our ability to recover high-redshift galaxies with NIRCam, we used a version of the JAGUAR mock catalog that was created without modeling Ly$\alpha$ but is otherwise identical. We repeated our color-cut analysis at the DEEP survey depth, with a two-color selection and a detection SNR of 3. We plot these results in Figure \ref{fig:LyAComparison}. 

The presence of Ly$\alpha$ emitted by a galaxy has a subtle effect on dropout selection. We can illustrate this by looking at the TPD and completeness for the F070W dropouts. At blue selection colors, these values are higher for the sample without Ly$\alpha$ emission, and then at redder selection colors they are higher for the sample that includes Ly$\alpha$ emission. For F070W dropouts, we select objects at $z > 4.11$, which includes objects where Ly$\alpha$ is entering the F070W band, enhancing the flux, and making the F070W-F090W color bluer than it would otherwise be without Ly$\alpha$ emission. At the same time, for objects at a redshift were Ly$\alpha$ sits in the F090W filter, this contributes to the flux in this band, causing these mock galaxies to be \textit{redder} in the Lyman break color cut, and \textit{bluer} in the UV continuum cut. Mock galaxies with Ly$\alpha$ emission are then both bluer and redder in the Lyman break cut depending on their redshifts, which impacts their selection as seen in the top and middle left panels of Figure \ref{fig:LyAComparison}. This effect is also observed for the TPD in the F090W and F115W dropout panels, but at less significance and redder selection colors. We find that the accuracy for dropout samples without Ly$\alpha$ emission is higher than samples with the emission line, with the highest significance for F115W dropouts. Similar results were seen for observations of galaxies in the \textit{Hubble} Ultra Deep Field (HUDF) with the Multi Unit Spectroscopic Explorer (MUSE) in \citet{inami2017}, where these authors present HST color cuts to select for Ly$\alpha$ emitters at $2.9 < z < 6.7$.

\subsection{Alternate Color Selection Criteria}
\label{sec:alternateselection}

Thus far, we have only explored NIRCam color selections using three adjacent photometric bands (along with non-detections in photometric bands shortward of the Lyman break). Lyman break selection, however, uses a pair of observed colors: one that spans the Lyman break at a particular redshift and one that covers the relatively featureless UV stellar continuum from massive stars. In this section, we examine the TPD, completeness, and accuracy for alternate UV continuum color cuts which utilize two unique photometric bands (``Four-band color selection'') and a scenario where the UV continuum cut attempts to span the entire rest-UV portion of the galaxy SED (``long UV baseline'').

In the three-band selection methods we have outlined thus far, mock galaxies can artificially be driven into or out of the selection boxes because of noise in the common photometric band. To help explore this effect, we also explored selecting Lyman break galaxies using photometry with four distinct NIRCam bands. While a four-band color selection criterion would require additional deep observations, it has the added benefit that noise in a single photometric band cannot affect both colors being used to select the galaxy. 

\begin{figure*}[ht!]
    \centering
    \includegraphics[width=0.85\textwidth]{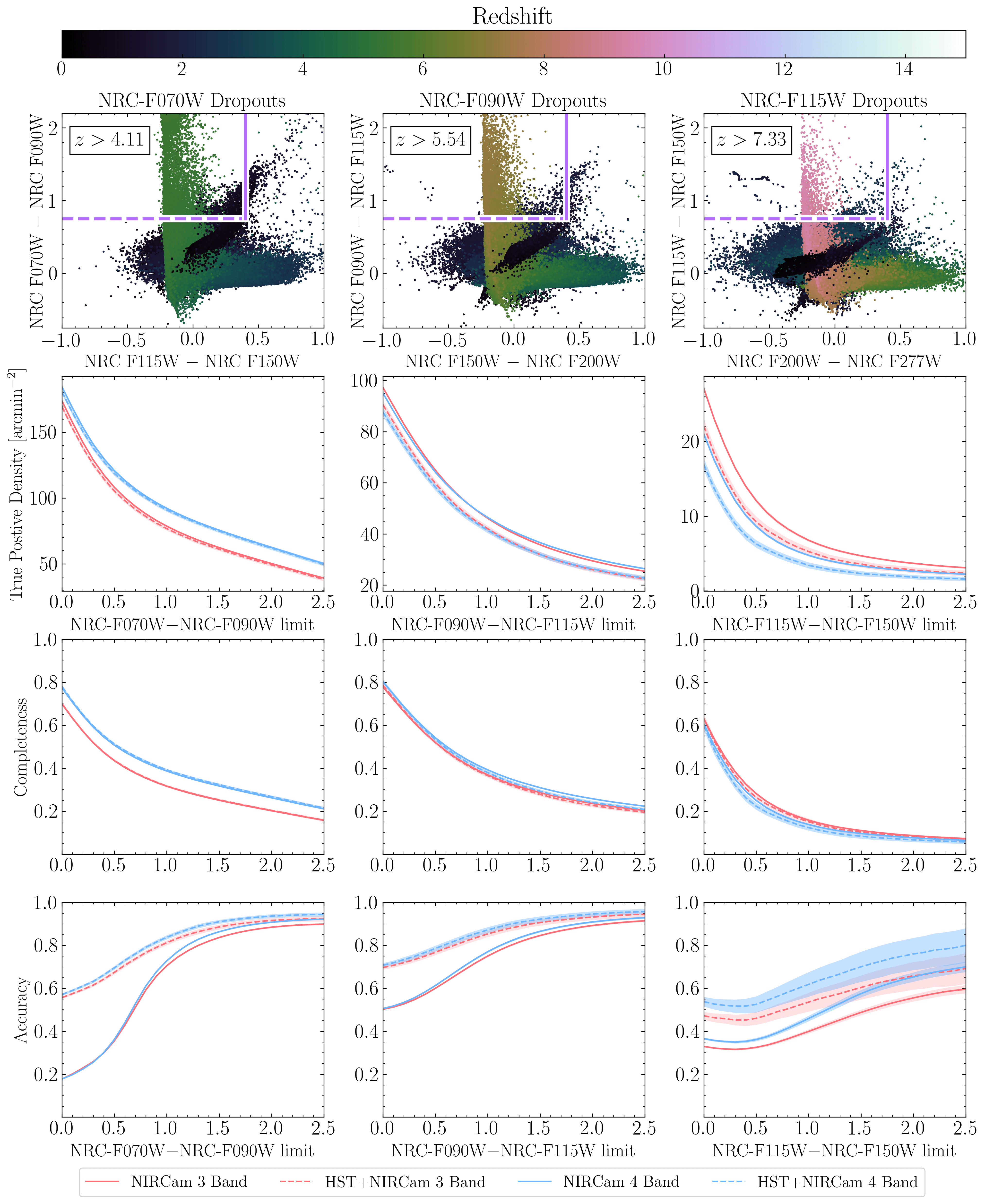}
	\caption{Color-color plots (first row) and the resulting TPD (second row), completeness (third row) and accuracy (fourth row) plots as a function of color cut for F070W dropouts (left), F090W dropouts (middle), and F115W dropouts (right) with three-band (red) and four-band (blue) selection for NIRCam (solid) and \textit{HST}+NIRCam (dashed) surveys. For comparison, we require a UV continuum color cut of $< 0.4$. TPD and completeness are higher with four-band selection than with three-band selection for F070W dropouts, similar between the two methods for F090W dropouts, and lower with four-band selection for F115W dropouts. With the exception of F115W dropouts, the accuracy of the four-band color selection is relatively consistent with that of the three-band selection. \label{fig:4BandsComparison}} 
    \end{figure*}

For the four-band analysis, we updated our selection criteria and re-ran the selection tests as was done in previous sections. For F070W dropouts, we compared F070W - F090W and F115W - F150W colors. For F090W dropouts, we compared F090W - F115W and F150W - F200W colors. Finally, for F115W dropouts, we compared F115W - F150W and F200W - F277W colors. In all cases, we used the DEEP survey depth, with a 3$\sigma$ detection, and explore the two-color cut selection, varying the Lyman break cut (we fixed the UV continuum cut in each case using a similar test to what was done in Section \ref{sec:surveydepth} for the three-band selection). In Figure \ref{fig:4BandsComparison}, we plot the TPD and accuracy for the four-band selection criteria compared to the three-band selection criteria.

For both F070W and (with less significance) F090W dropouts, the four-band color selection results in a larger TPD and completeness and a higher accuracy at all color cuts we explored. For three-band selection, dropout galaxies that are at redshifts where the Lyman break has entered one of the bands used in the UV color cut will be rejected for being too red, which results in fewer total high-redshift galaxies selected. With four-band color selection, the UV color cut samples a longer wavelength region of the SED, and this effect is not observed, leading to a higher TPD. For F115W dropouts, this effect is less significant (due to the declining number of very high-redshift galaxies), and the addition of the F277W SNR requirement results in lower TPD values, although this is also at an increased accuracy. As a result, our results demonstrate that four-band selection is recommended for F070W and F090W dropouts, or at very large areas (where the impact to the recovered TPD is minimal) for F115W dropouts.

    \begin{figure*}[ht!]
    \centering
    \includegraphics[width=0.85\textwidth]{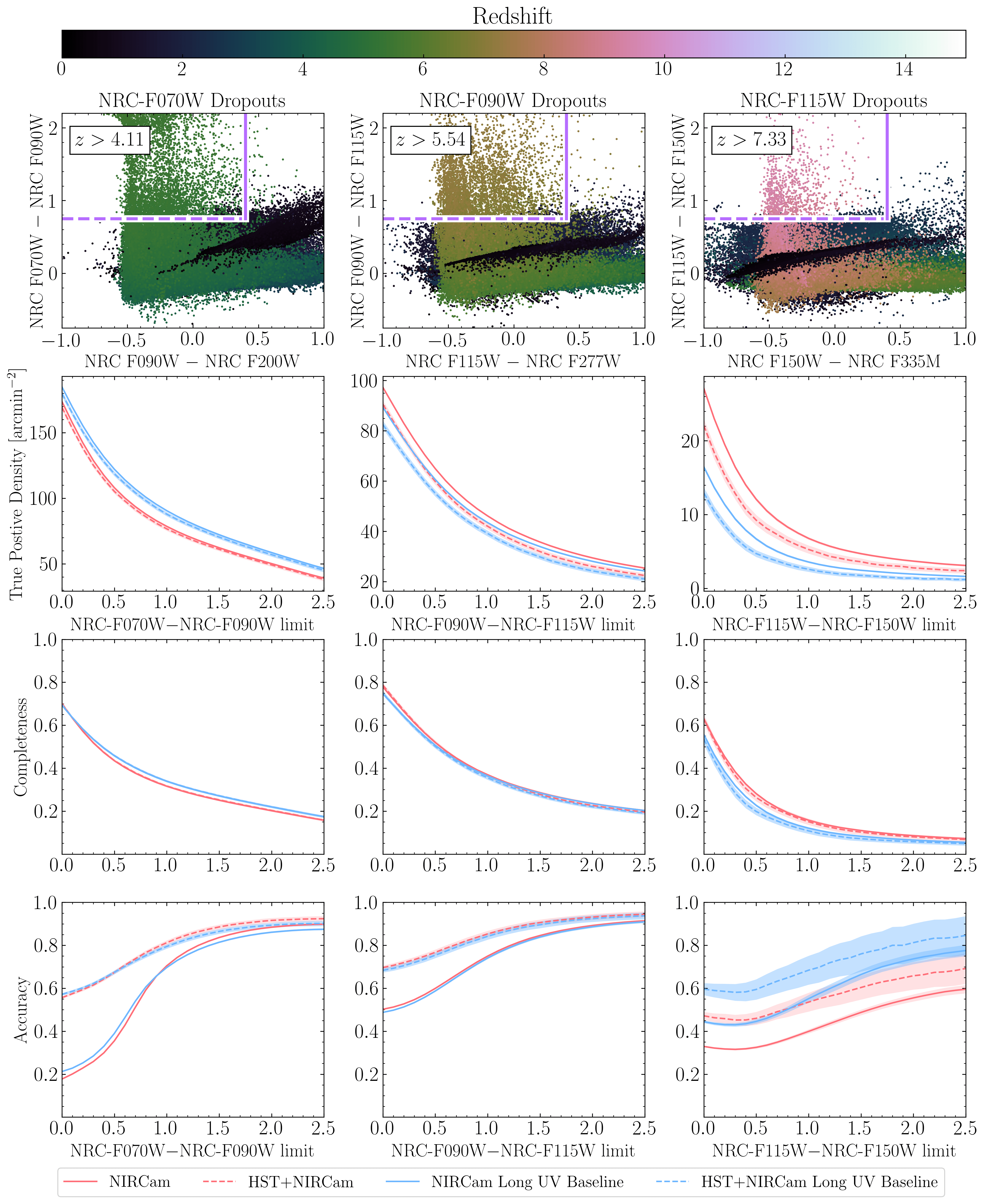}
	\caption{Color-color plots (first row) and the resulting TPD (second row), completeness (third row), and accuracy (fourth row) plots as a function of color cut for F070W dropouts (left), F090W dropouts (middle), and F115W dropouts (right), with the three-band selection from Section \ref{sec:surveydepth} (red) and with a three-band selection that utilizes a longer UV baseline (blue) for NIRCam (solid) and \textit{HST}+NIRCam (dashed) surveys. For comparison, we require a UV continuum color cut of $< 0.4$. \label{fig:LongUVBaselineComparison}} 
    \end{figure*}

    \begin{figure*}[ht!]
    \centering
    \includegraphics[width=0.85\textwidth]{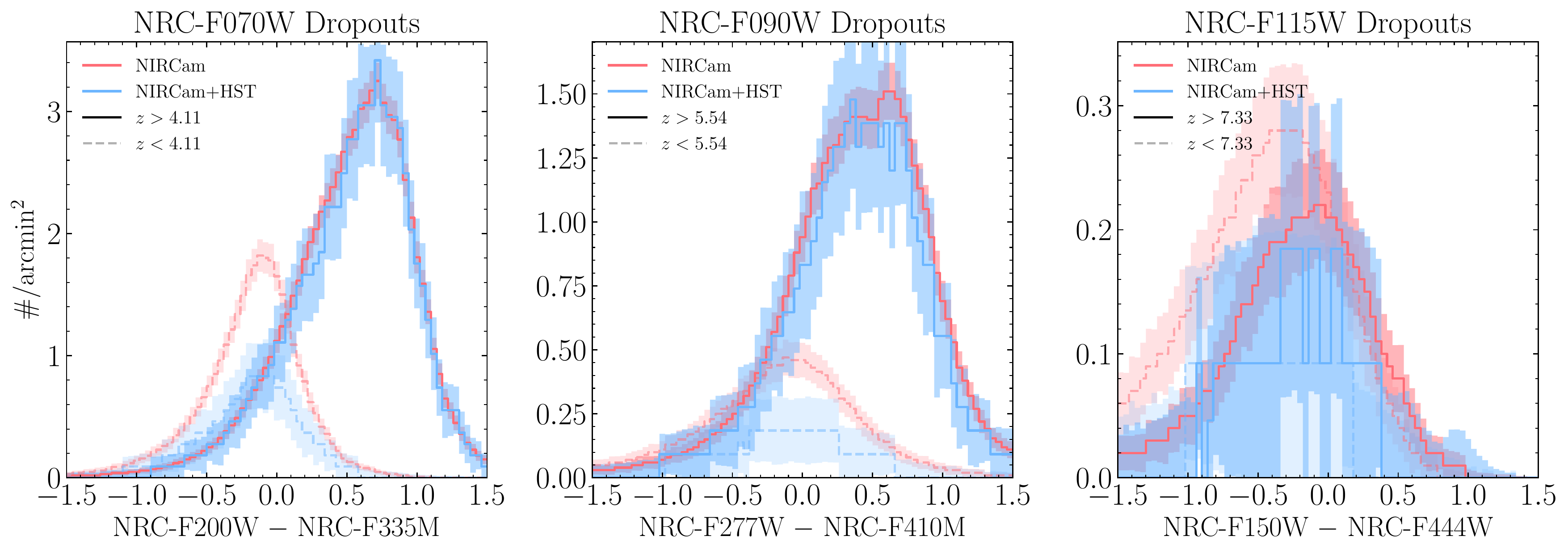}
    \caption{ \label{fig:redrejectionhistograms} Color histograms of mock galaxies at the DEEP survey depth with F070W - F090W $> 1.0$ (left), F090W - F115W $> 1.0$ (middle), and F115W - F150W $> 1.0$ (right), as well as a second color cut as described in Figure \ref{fig:DepthComparison}, with a detection SNR > 3 for all three plots. In the left panel, we plot F200W - F335M color, in the middle panel we plot F277W - F410M color, and in the right panel we plot F150W - F444W color. In each panel, true high-redshift objects are plotted with a solid line and lower-redshift interlopers are plotted with a dashed line. We show results from NIRCam-only photometry only with a red line, and HST+NIRCam with a blue line. For F070W and F090W dropouts, the true high-redshift objects are found at redder short-to-long-wavelength colors than the interlopers. For F115W dropouts, because NIRCam photometry does not cover the optically red portion of the SED in interloper galaxies, the difference between interloper and true high-redshift galaxy colors is less pronounced.} 
    \end{figure*}

We also examined a selection criteria where the UV continuum cut spans a longer wavelength range across the rest-frame ultraviolet. For F070W dropouts, in this alternate selection criteria, the UV continuum cut is F090W - F200W, for F090W dropouts, the alternate UV continuum cut is F115W - F277W, and for F115W dropouts, the alternate UV continuum cut is F150W - F335M. As with the four-band color selection, we updated our selection criteria and re-ran the selection tests for the DEEP survey depth and a detection SNR > 3.0, and explore the two-color cut selection, varying the Lyman break cut (we fixed the UV continuum cut in each case). In Figure \ref{fig:LongUVBaselineComparison}, we plot the updated color-color diagrams for F070W (left column), F090W (middle column), and F115W (right column) dropout selection with a longer UV baseline. Here, we compare the TPD (second row), completeness (third row), and accuracy (fourth row) between cuts estimated with this longer UV baseline color criteria to those made with the original criteria from Section \ref{sec:surveydepth}. For F070W dropouts, the use of a longer UV baseline results in a higher TPD and completeness but at similar accuracy values. This is likely because of the increased sensitivity for the F200W filter compared to the F115W filter, which leads to more objects satisfying the detection threshold. For F090W dropouts, the TPD and completeness is lower when using a longer UV baseline at similar (but slightly lower) accuracy values. For F115W dropouts, while the TPD decreases by a factor of two, the completeness only decreases slightly, but the accuracy increases when utilizing a longer UV baseline. These differences reflect how a longer UV baseline results in a larger spread in color values, as seen in the first row of Figure \ref{fig:LongUVBaselineComparison}, causing more objects to scatter outside of the selection boxes. 

\subsection{NIRCam Long Wavelength Rejection Colors}
\label{sec:dichroicrejection}

One of the key features of the NIRCam instrument is a dichroic beam splitter which allows observations in two filters simultaneously, one at short wavelengths (0.6 - 2.3 $\mu$m), and one at longer wavelengths (2.4 - 5.0 $\mu$m). Future deep extragalactic surveys with NIRCam will need to utilize the dichroic to increase the efficiency of the observational strategy. In this section, we will discuss the usage of NIRCam observations made at longer wavelengths to help select high-redshift galaxies (following work done in previous sections) as well as reject low-redshift interlopers. 

For Lyman break selection, in this current work we only explore photometric bands that cover the rest-frame UV, as at longer wavelengths the addition of flux due to strong emission lines and the 4000\AA\ +Balmer break results in redder UV-to-optical colors. In Section \ref{sec:alternateselection}, we demonstrate the TPD, completeness, and accuracy for a UV continuum color cut with a longer wavelength range. For F070W dropouts, the UV extends only to the F200W filter, and so longer wavelength data will only probe the rest-frame optical and near-IR. However, for F090W and F115W dropouts, observations can be made with the F277W and F335M filters respectively, which can be done simultaneously alongside shorter wavelength observations.

In addition, longer wavelength data can be used to reject low-redshift interloper galaxies by virtue of the overall color differences between short and longer wavelength observations between these two samples. In true high-redshift galaxies, NIRcam long-wavelength filters cover the rest-frame optical (see Figure \ref{fig:LyADropoutExample}), which may have boosted flux due to the 4000\AA\ +Balmer break and optical line emission. In interlopers, however, NIRCam long-wavelength data samples the continuum drop-off in the near-IR (in the absence of significant very hot dust emission). By comparing a short to a long wavelength filter, interloper mock galaxies in JAGUAR are observed to be systematically bluer than the true high-redshift mock galaxies. 
    
In Figure \ref{fig:redrejectionhistograms}, we plot color distributions for mock galaxies observed as part of our DEEP survey, with a detection SNR > 3.0. In each panel, we plot the distribution of dropout galaxies above our redshift cuts with solid lines, and the interloper galaxies in dashed lines, and we plot in red and blue the distributions with NIRCam data alone and NIRCam+HST data respectively. In the left panel, we plot the F200W - F335M color distribution, and the distribution of high-redshift dropout mock galaxies is significantly redder than the interloper distribution. For F070W dropout galaxies at $z > 4.11$, the NIRCam F335M filter covers the [OIII]$\lambda5007$ emission line, leading to the red color. The [OIII]$\lambda$5007 emission line has been inferred to be strong (with EW values of $500 \AA$) at these redshifts from Spitzer IRAC observations \citep{labbe2013, smit2015, debarros2019}. In the middle panel, we plot the F277W - F410M color for F090W dropouts and interlopers, and we see a similar behavior, as the F410M filter covers [OIII]$\lambda5007$ for true high-redshift dropout galaxies. The difference is not as great as what is observed for the F070W dropouts, as the F410M filter no longer samples the near-IR wavelength range for the interloper galaxies. We find that a color cut at F200W - F335M > 0.0 or F277W - F410M > 0.0 aids in rejecting interloper galaxies.

The situation for F115W dropouts is more complex because NIRCam short-to-long-wavelength colors are very similar in both true high-redshift galaxies and interlopers. In the right panel of Figure \ref{fig:redrejectionhistograms}, we plot the F150W - F444W (the longest wavelength wide-band NIRCam filter) colors for F115W dropout galaxies at $z > 7.33$ and interloper galaxies. We see that there is a tendency for interlopers to be at slightly bluer colors than true high-redshift mock galaxies, although with less significance owing to the small numbers of these galaxies in a given sample. We explored other color combinations besides F150W - F444W, but each had similar or worse results for rejecting interlopers.

    \begin{figure*}[ht!]
    \centering
    \includegraphics[width=0.85\textwidth]{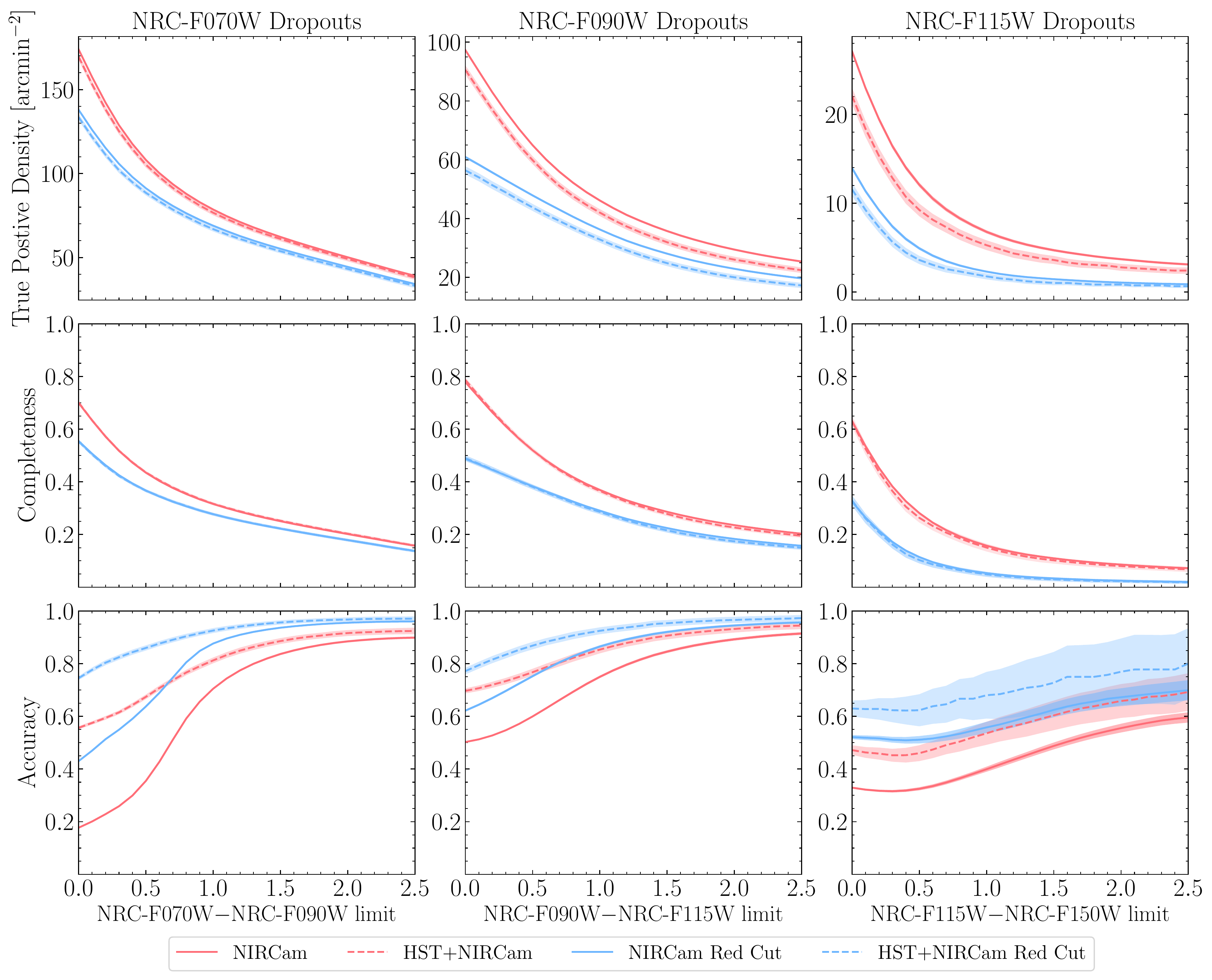}
    \caption{\label{fig:redrejectionTPDAccuracy} TPD (top row), completeness (middle row), and accuracy (bottom row) as a function of color cut for F070W dropouts (left), F090W dropouts (middle), and F115W dropouts (right), with (blue) and without (red) an additional short-to-long wavelength color cut for NIRCam (solid) and \textit{HST}+NIRCam (dashed) surveys. For F070W dropouts, we require F200W - F335M > 0.0, for F090W dropouts, we require F277W - F410M > 0.0, and for F115W dropouts, we require F150W - F444W > 0.0. We additionally require UV continuum color cut of F090W - F115W $< 0.4$ (left), F115W - F150W $< 0.4$ (middle), and F150W - F200W $< 0.4$ (right). While the addition of these short-to-long-wavelength color cuts results in a decrease in the TPD and completeness, the accuracy increases significantly for all three selection criteria.} 
    \end{figure*}

To explore the use of these cuts for F070W, F090W, and F115W dropout selection, we calculated the TPD, completeness, and accuracy values for mock galaxies at the DEEP survey depth with SNR > 3.0, but required F200W - F335M > 0.0 for F070W dropouts, F277W - F410M > 0.0 for F090W dropouts, and F150W - F444W > 0.0 for the F115W dropouts, which we plot compared to the TPD, completeness and accuracy made without the cuts in Figure \ref{fig:redrejectionTPDAccuracy}. The bottom panels show the increase in accuracy that can be achieved through these short-to-long wavelength color cuts, although this is at the expense of the TPD and completeness plotted in the top and middle panels, which both drop by almost half for F115W dropouts.

This sets up a potential example observational strategy for F070W, F090W, and F115W dropouts. For F070W dropouts, short-wavelength observations would need to be made at F070W, F090W, and F115W (or either F150W or F200W) for the dropout selection, but these data could be supplemented by simultaneous observations with F335M, as well as longer-wavelength data (F356W and F410M), which is important for any potential SED fitting of these galaxies. For F090W dropouts, it is much more straightforward. The short wavelength data necessary would be at F090W and F115W, which could be observed simultaneously with the F277W and F410M filters. Similarly, for F115W dropouts, the short wavelength data necessary would be at F115W and F150W, which could be observed simultaneously with the F335M (or F277W) and F444W filters. 

\subsection{Interlopers and $\chi^2_{\mathrm{opt}}$}
\label{sec:chisquareopt}

In \citet{bouwens2015}, the authors explore the usage of a statistic they refer to as $\chi^2_{\mathrm{opt}}$, defined as 
\begin{equation}
\chi^2_{\mathrm{opt}} = \Sigma_i \mathrm{SGN}(f_i)(f_i / \sigma_i)^2
\end{equation}

where for each undetected (SNR < 2) photometric band to the blue of the Lyman break, $f_i$ is the flux in that band, $\sigma_i$ is the uncertainty in that band, and $\mathrm{SGN}(f_i)$ is 1 if $f_i > 0$ and -1 if $f_i < 0$. This statistic was designed to measure whether, for objects with a non-significant detection in the filters to the blue of the Lyman break, the flux is biased towards positive values. For actual high-redshift galaxies, the distribution of $\chi^2_{\mathrm{opt}}$ should be centered at 0, while for lower-redshift interlopers, the distribution will be biased towards positive values. 

    \begin{figure*}[ht!]
    \centering
    \includegraphics[width=0.66\textwidth]{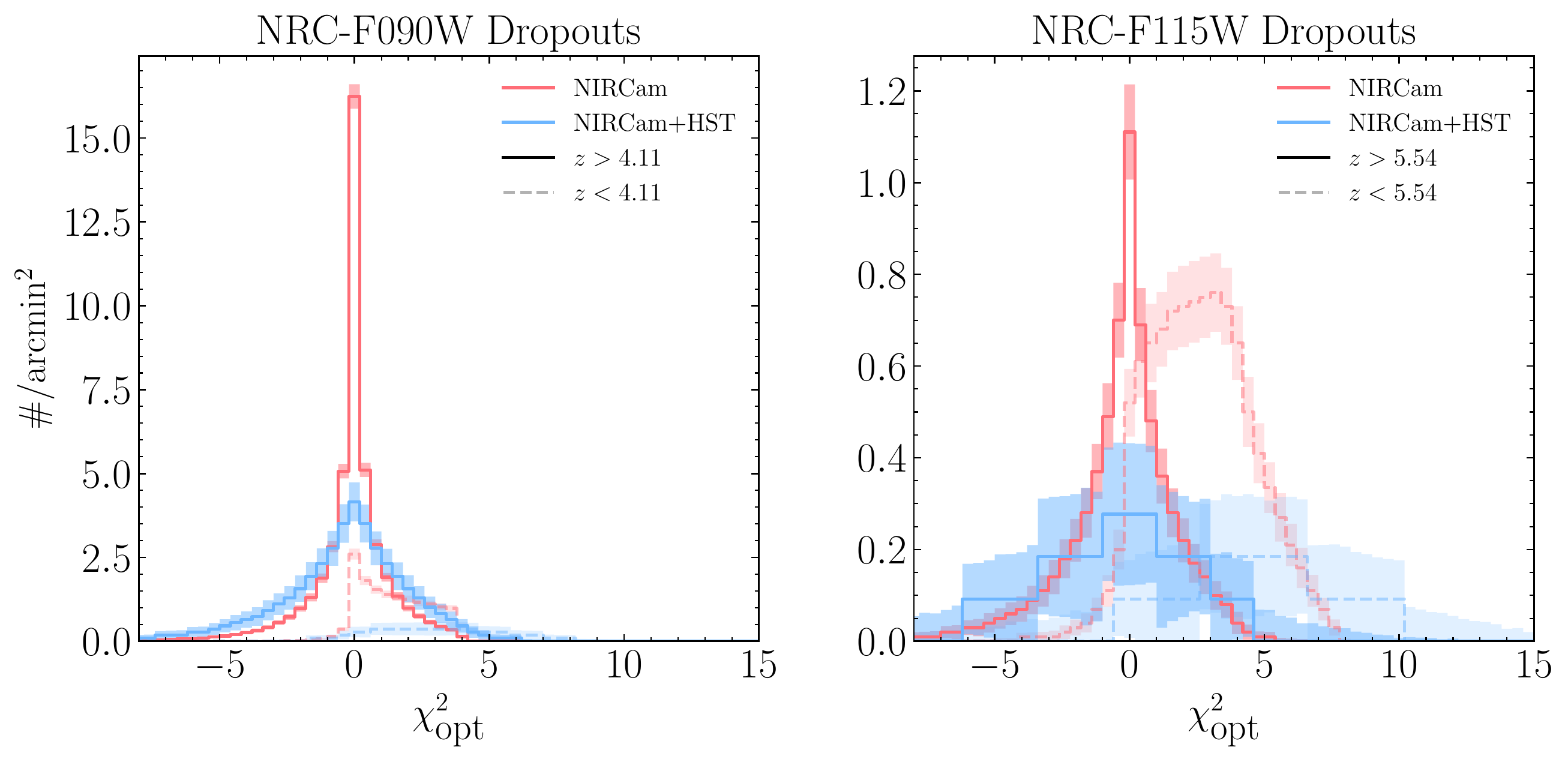}
    \caption{\label{fig:ChiOptComparison}  $\chi^2_{\mathrm{opt}}$ distributions for F090W dropouts (left), F115W dropouts (middle), and F150W dropouts (right) for high-redshift objects (solid) and low redshift interlopers (dashed) in NIRCam (red) and \textit{HST}+NIRCam (blue) surveys. The use of $\chi^2_{\mathrm{opt}}$ is more effective for higher redshift dropout galaxies where there are more photometric bands for a given mock galaxy at wavelengths shorter than the Lyman break. For the reddest dropout bands, a $\chi^2_{\mathrm{opt}} \gtrsim 5$ would accurately reject outliers without impacting the selection of true high redshift galaxies.} 
    \end{figure*}

    \begin{figure*}[ht!]
    \centering
    \includegraphics[width=\textwidth]{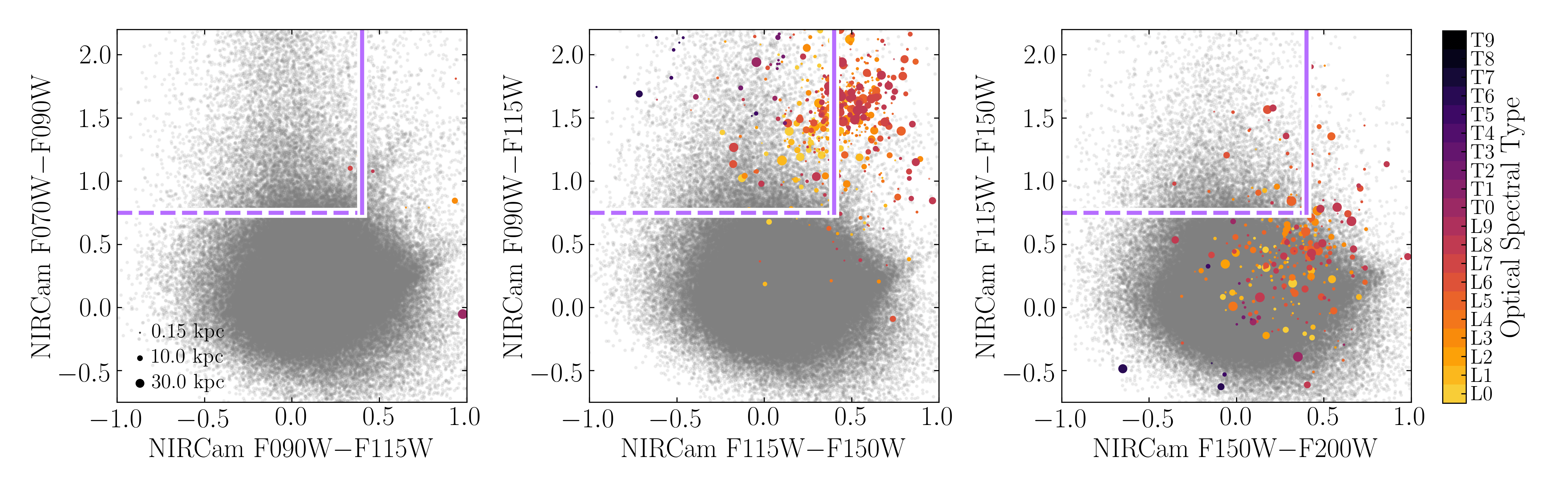}
    \caption{
    \label{fig:browndwarfcolors} NIRCam color-color plots with the mock galaxies from Figure \ref{fig:jaguarnoisycolor} plotted with grey points, overplotted with a selection of brown dwarfs, at different simulated distances and with at the DEEP \textit{HST}+NIRCam survey depth. The points are colored by their optical spectral type, as shown on the color bar, and the sizes of the points indicate the distance from Earth, as shown in the bottom-left corner of the leftmost figure.} 
    \end{figure*}

We explore the efficacy of the \citet{bouwens2015} $\chi^2_{\mathrm{opt}}$ statistic in discriminating low-redshift interlopers in \textit{JWST} surveys using our three-band analysis at the DEEP survey depth. We calculate the $\chi^2_{\mathrm{opt}}$ for each object selected with the two color cut adopted throughout this work assuming a detection SNR $>3$ and a blue non-detection SNR $<2$. We separate them based on their true redshifts and measure the distribution of the results, presented in Figure \ref{fig:ChiOptComparison}. We note that since there are no blue rejection filters for F070W dropouts, their $\chi^2_{\mathrm{opt}}$ cannot be calculated. With F090W dropouts, there are too few blue bands to find a clear delineation between high and low redshift mock galaxies. However, for the redder F115W dropout band, 10 - 30\% of interloper galaxies could be reliably rejected without affecting the number of high redshift galaxies selected by adopting a $\chi^2_{\mathrm{opt}} \gtrsim 5$. This cutoff is relatively unaffected by an increase the detection SNR.

\subsection{Brown Dwarf Interlopers}
\label{sec:browndwarfs}

In addition to the mock galaxies in the JAGUAR catalog, we also explored how ultracool brown dwarf stars may be selected as dropout candidates following the work of \citet{wilkins2014}, \citet{finkelstein2015}, and \citet{ryan2016}. Brown dwarfs have stellar spectra that become redder at cooler temperatures, with stronger molecular absorption features that can mimic the red dropout colors of high-redshift galaxies. While these studies conclude that ultracool dwarfs will be relatively rare ($\sim$1 arcmin$^{-2}$), extended deep \textit{JWST} surveys will likely contain a number of dwarfs due to the 9.7 arcmin$^{2}$ FOV of NIRCam. To that end, we used a subsample of the published spectra for L and T dwarfs from the SpeX Prism Spectral Library\footnote{Compiled by Adam Burgasser and found online at http://pono.ucsd.edu/~adam/browndwarfs/spexprism/} and calculated the fluxes of these objects through the NIRCam wide filters: F070W, F090W, F115W, F150W, and F200W, as these spectra have wavelength coverage to 2.5$\mu$m. We supplemented these observational data with a set of L and T dwarf model spectra (which extend to 50$\mu$m) from Sonora18 (Marley et al. in prep) at a range of surface temperatures (T = 200 - 2300 K) and a fixed surface gravity of $\log{(g)} = 5.0$. For these model spectra, we calculated the NIRCam fluxes through the NIRCam wide filters F070W, F090W, F115W, F150W, and F200W, F277W, F356W, and F410M and the NIRCam medium filters F335M and F410M. We simulate these real and model objects at a range of distances between 0.1 kpc and 40 kpc, and add noise at the DEEP \textit{HST}+NIRCam survey depth. We note that brown dwarfs are unresolved in extragalactic surveys and a stellarity parameter has been used to remove these sources from deep HST catalogs \citep[see Section 3.5.1. in][]{bouwens2015}. We do not simulate this in our current work, and caution that while morphology can be used for rejecting stellar contaminants, compact high-redshift galaxies may also be similarly unresolved. 

We plot the positions of the noisy brown dwarf candidates on the F070W, F090W, and F115W dropout color-color plots in Figure \ref{fig:browndwarfcolors}, where we impose the same red filter detection SNR (>3) and blue filter non-detection SNR (<2). In the figure, the colors of the points indicate the optical spectral type of the object given in the SpeX Library (as shown in the color bar on the right) or estimated from the temperature of the Sonora18 model spectrum, and the size of the point indicates the simulated distance of the brown dwarf. While brown dwarfs do not have colors similar to F070W dropouts, a large population of brown dwarfs would be selected as F090W dropouts, and a smaller number would be identified as F115W dropouts. 

    \begin{figure}[t!]
    \centering
    \includegraphics[width=\columnwidth]{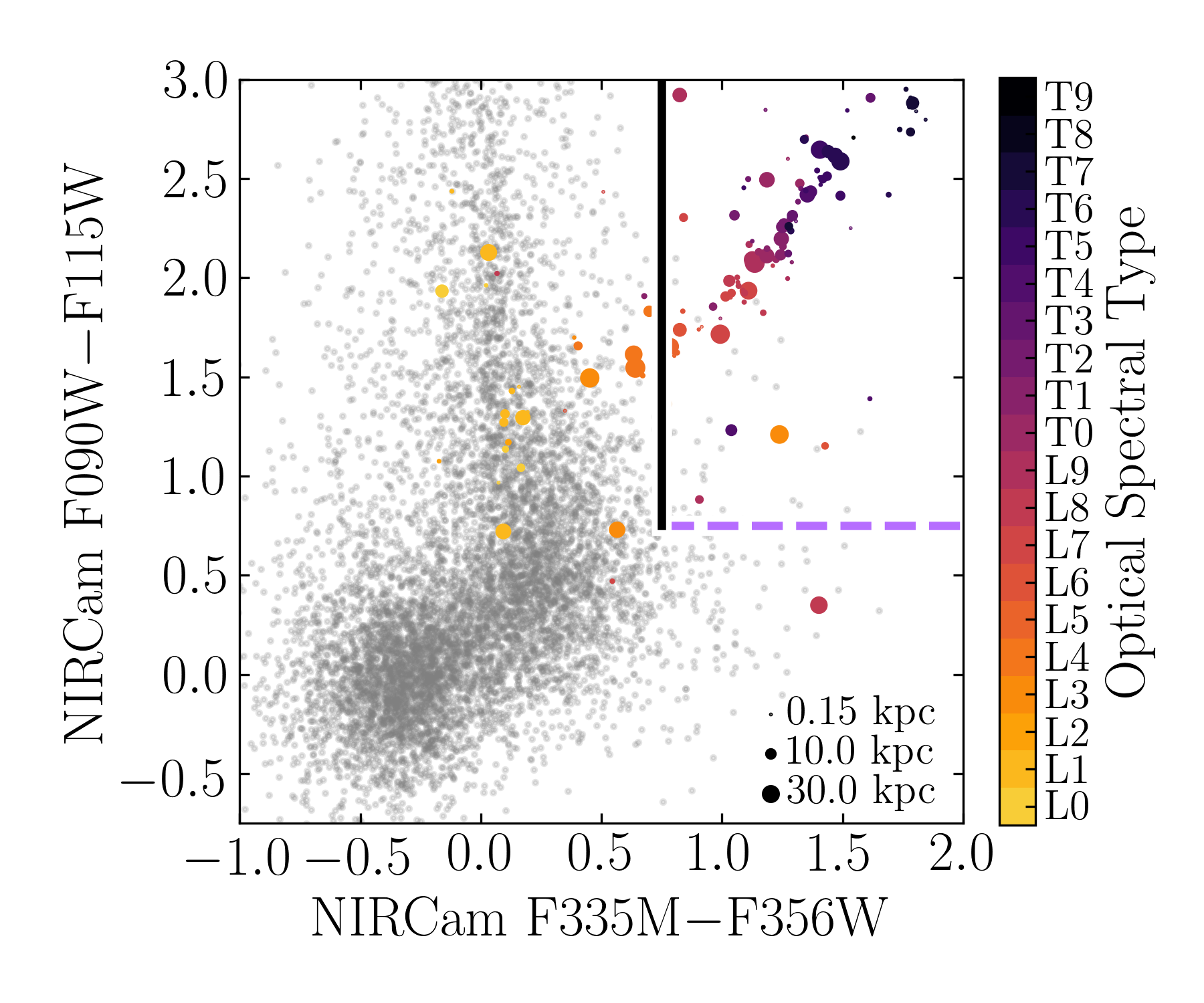}
    \caption{
    \label{fig:browndwarflongwavecolors} NIRCam F090W - F115W color vs. F335M - F356W color plot. In grey, we plot JAGUAR mock galaxies in a DEEP survey with F115W, F150W SNR > 3 and F435W, F606W, and F070W SNR < 2 at $z > 5.54$, the redshift demarcation for F090W dropouts. The colored points are Sonora18 model brown dwarfs with optical spectral type as shown with the color bar on the right side, with the size of the markers symbolizing the distance from the Earth as given in the bottom right of the plot. While the central panel in Figure \ref{fig:browndwarfcolors} shows that a quantity of L and T dwarfs contaminate F090W dropout selection, many of the late L and T dwarfs that satisfy the same F090W - F115W color cut (dashed lavender line) can be removed by also requiring F335M - F356W < 0.75 (black line).} 
    \end{figure}

For the F090W dropouts there is a general trend between optical spectral type and redder F090W - F115W and F115W - F150W colors, and most of the sources selected are at larger distances (> 10 kpc), echoing results from \citet{ryan2016} demonstrating that \textit{JWST} will be able to detect brown dwarfs in the Milky Way halo. A color selection at F115W - F150W < 0.3 would select against many late L and T dwarfs. Late T-dwarfs have very red F090W - F115W colors and blue F115W - F150W colors, and would also be selected as F090W dropouts. To aid in differentiating true high-redshift galaxies from brown dwarfs, we used the \textit{Spitzer} IRAC photometry for a sample of 86 late M, L, and T dwarfs provided by \citet{patten2006}. After converting the Channel 1 (3.6 $\mu$m) and Channel 2 (4.5 $\mu$m) fluxes to AB magnitudes, we find that M and L dwarfs have [3.6] - [4.5] $< -0.3$ (roughly analogous to NIRCam F356W - F444W < -0.3), which is significantly bluer than the bulk of true F090W dropout galaxies. T dwarfs in the \citeauthor{patten2006} sample, however, have red [3.6] - [4.5] colors and are not as easily separated from F090W dropouts. To find methods for removing T dwarfs from F090W dropout samples, we looked at the long-wavelength NIRCam colors of these stars using the Sonora18 model spectra. In Figure \ref{fig:browndwarflongwavecolors}, we plot the F090W - F115W color vs. F335M - F356W color for both JAGUAR mock galaxies and Sonora18 model brown dwarfs with F115W, F150W SNR > 3 and F435W, F606W, and F070W SNR < 2 (\textit{HST}+NIRCam) with the model brown dwarf points colored as they are in Figure \ref{fig:browndwarfcolors}. From this Figure, we show that Late L and all T dwarfs can be reliably separated from true F090W dropouts by requiring a color cut at F335M - F356W < 0.75 (black vertical line), along with the F090W - F115W color cut (lavender horizontal dashed line), although this potentially removes a small population of high-redshift mock galaxies with strong optical line emission. 

    \begin{figure*}[ht!]
    \centering
    \includegraphics[width=\textwidth]{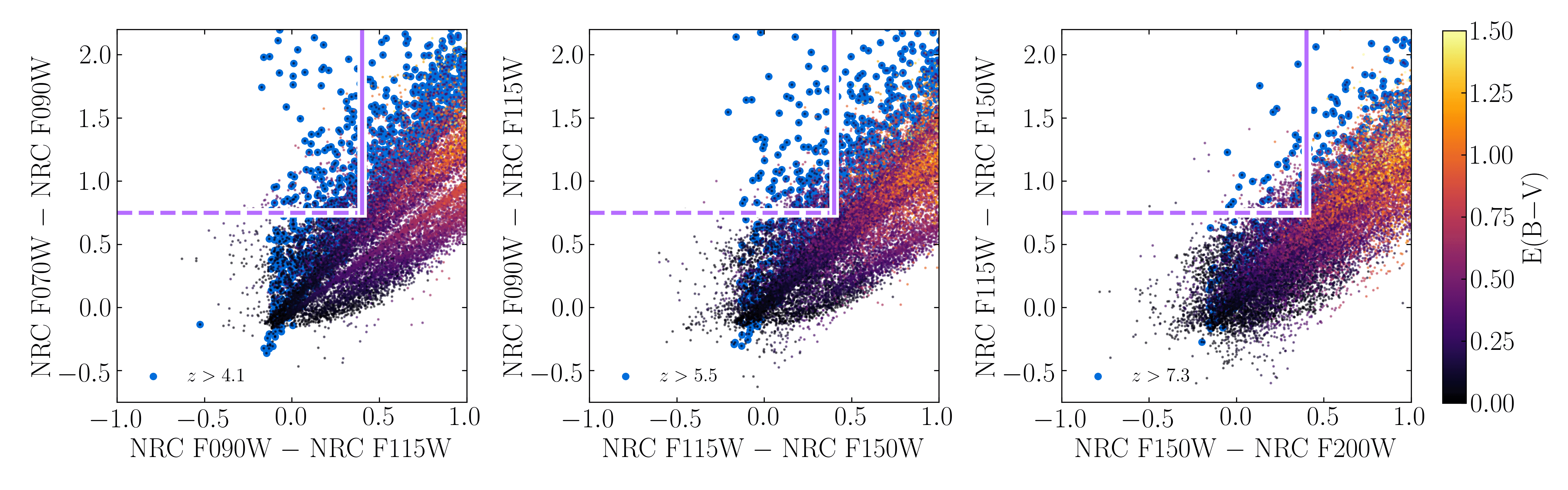}
    \caption{
    \label{fig:dustygalaxycolors} NIRCam color-color plots as in Figure \ref{fig:jaguarcolor}, but with mock galaxies with randomly assigned values for the extinction between $E(B-V) = 0 - 2$. For mock galaxies at $z < 3$, we use the \citet{calzetti2000} dust prescription, and at $z > 3$, we use the SMC-bar-like dust from \citet{gordon2003}. We represent actual high-redshift dropout mock galaxies in each panel with blue points. Dust causes the mock galaxy colors to be redder along each axis of the figure, reinforcing the need for a second color cut (such as the example vertical solid lavender line in each panel) to remove dusty low-redshift interlopers.} 
    \end{figure*}

It is important to note that we do not simulate the on-sky density of objects in our sample in these plots. \citet{ryan2016} explored the actual number density of brown dwarfs in the thick and thin disk of the Milky Way, and concluded that only a few ultracool dwarfs would contaminate extragalactic surveys. Our results demonstrate the importance in the use of a Two Color or Compound Color method as well as observations at longer wavelengths to help mitigate this contamination. 

\subsection{Dusty Star-Forming Galaxies}
\label{sec:dustygalaxies}

The JAGUAR catalog contains only a limited population of highly dust obscured star-forming mock galaxies as the mass and luminosity functions that were used to create JAGUAR are dependent on observations of the rest-frame optical and UV portions of a galaxy's spectrum and are therefore missing extremely dusty galaxies. To further explore how dust affects dropout selection, we reproduced the entire catalog of JAGUAR star-forming mock galaxies, keeping the properties including mass, observed redshift, and star formation history the same, but assigning a random extinction value (parameterized by the color difference E(B-V)) to each object between $E(B-V) = 0 - 2$. To recreate these objects, we used the Flexible Stellar Population Synthesis code \citep[FSPS,][]{conroy2009,conroy2010}, and used Padova isochrones along with the MILES spectral library \citep{miles2006}. We chose to model the dust as a foreground screen using the \citet{calzetti2000} prescription at $z < 3$, and with SMC-bar-like dust \citep{gordon2003} at $z > 3$. While assigning a random E(B-V) to all of the JAGUAR mock galaxies is nonphysical in light of observed trends between stellar mass and E(B-V) out to $z \sim 6$ \citep{schaerer2010}, as well as the complexity of actual dust geometry within galaxies, these extreme values for dust extinction will allow us to observe how obscuration affects the NIRCam colors of a diverse population of low-redshift interlopers.

In Figure \ref{fig:dustygalaxycolors}, we plot the noise-free F070W, F090W, and F115W dropout space of the dusty star-forming mock galaxies, with the points colored by E(B-V), as shown in the color bar on the right side of the figure. We indicate true high-redshift mock galaxies by blue circles. In each panel, dustier mock galaxies are found in a sequence that extends upwards and to the right. We also plot example two-color selection boxes, demonstrating the use of a second color cut to exclude lower redshift dusty interlopers. The lack of dusty mock galaxies that fall inside the selection boxes demonstrates the bias against selecting obscured galaxies using the Lyman dropout technique.  

\section{Comparison to the Empirical Galaxy Generator}
\label{sec:EGG}

In anticipation of future deep extragalactic surveys, the ASTRODEEP collaboration developed the Empirical Galaxy Generator \citep[EGG,][]{schreiber2017}\footnote{https://cschreib.github.io/egg/}, which constructs mock catalogs including both photometry and morphologies. Similar to JAGUAR, EGG uses empirical prescriptions, starting with a derivation of the evolution of the stellar mass function from deep observations. In this section we compare the recovered TPD, completeness, and accuracy for the EGG catalog to what we found using the JAGUAR catalog. A few of the primary differences between JAGUAR and EGG that will influence the present analysis are the evolution of the stellar mass function, the treatment of galaxy morphologies and dust obscuration, and the inclusion of self-consistent nebular continuum and line emission.

The EGG team started with a framework for the evolution of the star-forming and quiescent galaxy mass function at $z = 0.3 - 4.5$ based on observations from CANDELS \citep{grogin2011, koekemoer2011}, where they computed photometric redshifts using EAZY \citep{brammer2008} and galaxy stellar masses using FAST \citep{kriek2009}. At $z = 4.5 - 7.5$, the authors rely on the stellar mass functions from \citet{grazian2015}. The resulting mass function evolution has a steeper low-mass slope than the prescription that underpins the JAGUAR catalog at $z > 1.5$ and the discrepancy is larger at higher redshifts. In addition, the evolution of the EGG mass function predicts fewer high-mass galaxies at $z > 4$ than JAGUAR. Both of these differences are likely a consequence of the necessary extrapolation that was done for each catalog due to lack of observational data.

The SEDs in EGG were generated by first assigning a $U - V$ and $V-J$ color to each mock galaxy based on the observed evolution of these colors for star-forming and quiescent galaxies. At this point, each mock galaxy was given an SED based on the average SED for observed CANDELS galaxies with those $UVJ$ colors (from the FAST fits, using the \citet{bc2003} stellar library). As the morphology of each EGG mock galaxy is defined to be a combination of a bulge and disk component, each component was assigned a separate SED. This process differs significantly from the SED creation in JAGUAR, which uses BEAGLE fits to 3D-HST objects to calculate the SEDs for each object. In the version of the EGG catalog generation tool we used in this analysis, v1.4.0 (\textit{egg-gencat}), the authors included a simple prescription for emission lines, where the strength of each line is estimated using each mock galaxy's SFR, metallicity, total infrared luminosity, and gas mass\footnote{see: https://github.com/cschreib/egg/blob/master/CHANGELOG}, which we include to better compare to JAGUAR.

We used \textit{egg-gencat} to create two catalogs, one with 100 square arcminutes and one with 10.8 square arcminutes, with a minimum stellar mass of 10$^6$ M$_{\sun}$, at $z = 0.2 - 15$. We then constructed 500 noisy catalogs with each area in the exact manner as was done in Section \ref{sec:NIRCamNoise} for the JAGUAR catalogs, although we modified this process to account for the combination of the disk and bulge components in each EGG mock galaxy. From these noisy catalogs, we measured the TPD, completeness, and accuracy as a function of color cuts, SNR, and survey depth following the analysis we performed for the noisy JAGUAR catalogs.

\begin{figure*}[ht!]
    \centering
    \includegraphics[width=0.85\textwidth]{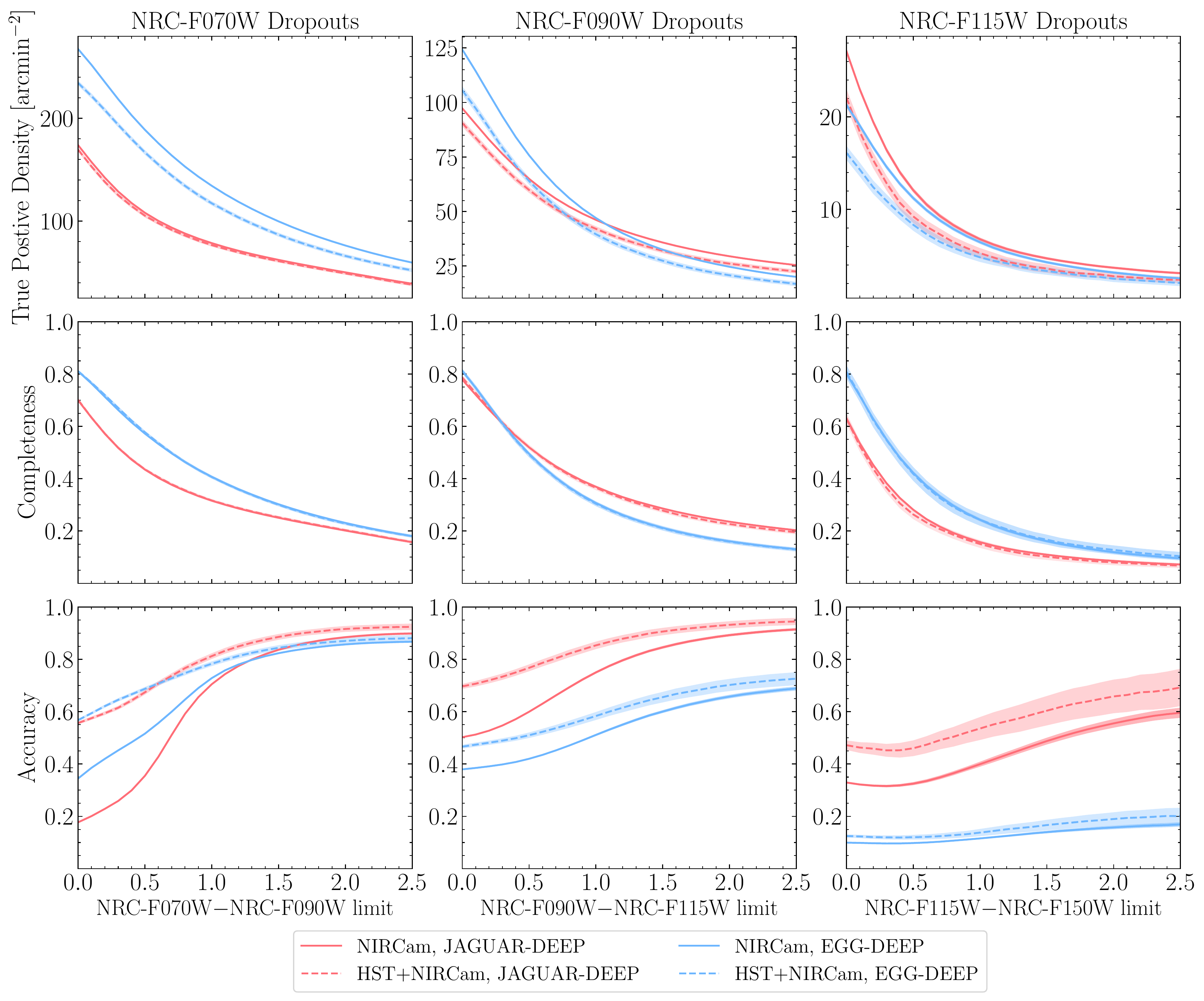}
	\caption{
	\label{fig:eggcomparison} TPD (top row) and accuracy (bottom row) as a function of color cut for F070W dropouts (left), F090W dropouts (middle), and F115W dropouts (right) in the JAGUAR (red) and EGG (blue) catalogs for NIRCam (solid) and \textit{HST}+NIRCam (dashed) surveys. In all three cases, we are using the DEEP survey depth, with a detection SNR > 3, and a two-color selection, with a UV continuum color cut of $< 0.4$. The evolution of the mass function in the EGG catalog predicts a much larger number of low-mass galaxies than the JAGUAR galaxies, and less high-redshift galaxies at all masses. This leads to a higher TPD for F070W and F090W dropouts, along with a much lower accuracy for F090W and F115W dropouts, where more faint low mass galaxies serve as high-redshift interlopers.} 
    \end{figure*}

In Figure \ref{fig:eggcomparison}, we plot the TPD, completeness, and accuracy as a function of Lyman break color cut at the DEEP survey depth, with a detection SNR > 3, and the Two Color Cut Scheme where the UV continuum cut is set at 0.4 for all three selection methods. We compare the EGG values to the JAGUAR values in each panel. For F070W dropouts, we find that EGG predicts almost double the TPD, and increased completeness, at all colors, but at significantly reduced accuracy, likely a result of the increased number of low-mass, faint galaxies in the EGG catalog. For F090W dropouts, the predicted TPD is more comparable between the EGG and the JAGUAR results, although the abundance of low-mass mock galaxies means that the predicted accuracy is significantly lower for EGG than for JAGUAR. For F115W dropouts, the EGG catalog results in higher completeness values but at very low accuracies. 

These results highlight the difficulty in extrapolating the galaxy stellar mass function to low mass, especially at high redshift. Future deep \textit{JWST} surveys will include a significant number of low-mass ($<10^7$M$_{\sun}$, see Figure \ref{fig:specmasshistograms}) galaxies that are currently too faint to be observed with existing instruments, which will allow for a more robust measurement of the low-mass slope of the mass function. 

\section{Discussion}
\label{sec:discussion}

\textit{JWST}/NIRCam will be an exceptional instrument for selecting large samples of high-redshift galaxies. The NIRCam field of view spans 9.7 arcmin$^2$, so with a single pointing as part of a DEEP survey, these results indicate that it may be possible to select $\sim900$ $z = 5 - 7$ F070W dropouts at greater than 70\% accuracy, $\sim500$ $z = 6 - 9$ F090W dropouts at greater than 70\% accuracy, and $\sim60$ $z = 8 - 11$ F115W dropouts at greater than 50\% accuracy. In Figure \ref{fig:specmasshistograms}, we additionally show that these samples include mock galaxies at stellar masses of 10$^6$ - 10$^7$ M$_{\sun}$, a stellar mass range that is difficult to access at these redshifts. 

These results also demonstrate the importance of targeting fields with existing deep \textit{HST} data. The addition of \textit{HST} photometry has a limited effect in reducing the number of recovered high-redshift galaxies, but serves to significantly increase dropout selection accuracy, especially for blue color limits. While the majority of existing deep \textit{HST} data are not at the XDF depths, the usage of shallower photometry should produce a result that is bracketed by the NIRCam only and \textit{HST}+NIRCam results we have presented. Additional exploration of alternate survey designs at different HST and NIRCam depths can be accomplished using the NIRCPrepareMock software package described in Section \ref{sec:NIRCamNoise}. 

A selection scheme should be chosen based on the sample purity and accuracy requirements for the science goal of a given survey. For F070W, F090W, and F115W dropouts, with two color selections, we find that a color cut of $\sim 1$ balances the trade-off between completeness (and TPD) and accuracy. Because of how the presence of significant quantities of dust affects the observed NIRCam colors of low redshift interloper galaxies, we also recommend additional color cuts to focus on galaxies with blue UV slopes. For more inclusive (bluer) color cuts, the Compound Color Cut method provides the highest accuracy levels observed, cutting out a quantity of low redshift interloper galaxies. In addition, when existing deep optical data exist, we recommend the use of the $\chi^2_{\mathrm{opt}}$ statistic to help increase survey accuracy, but with the caveat that there is significant overlap between the $\chi^2_{\mathrm{opt}}$ distributions for true high-redshift galaxies and lower redshift interlopers. 

\section{Conclusions}
\label{sec:conclusions}

We have demonstrated how \textit{JWST}/NIRCam colors can be used to select samples of high-redshift galaxies through the Lyman dropout technique by simulating surveys with the JAGUAR catalog. We examined how the exact color cut affected the on-sky TPD of high-redshift sources, and the completeness and accuracy of the resulting sample. Our primary results are:

\begin{enumerate}
    \item At increasingly redder color cuts, in all cases, the true positive density and completeness of the recovered sample decreases, as fewer objects are selected, but the overall accuracy increases, often plateauing at a maximum accuracy level for a given set of mock galaxies. This process is primarily driven by SNR, as the recovered mock galaxies in each selection method we explored are at higher redshifts than the interlopers, which are scattered into the selection box due to noise.
    \item We explored three dropout selection methods which utilize either one single color cut, two color cuts, or a compound method with three color cuts. While the use of a second or third color cut significantly increases the accuracy of the recovered sample, it is at the expense of the TPD and completeness, especially for F070W dropouts. This is less true for F115W dropout selection due to the limited number of mock galaxies predicted to be observed per square arcminute in a given survey. Dust obscuration has the effect of moving mock galaxies to redder colors, reinforcing the usage of two or three color selection criteria for selecting against low-redshift dusty interlopers.
    \item We find that the presence of Ly$\alpha$ emission has a complicated effect on the TPD, completeness, and accuracy of recovered samples as a function of color cut due to how the emission line contributes to flux in different bands at different redshifts. For F090W and F115W dropouts, we predict a higher TPD, completeness, and accuracy for samples without Ly$\alpha$ emission at most moderate color selection cuts. 
    \item Filter selection for the UV continuum color cut plays a significant role in recovering galaxies. For F070W dropouts, it is recommended to use filters that probe longer wavelengths to increase the TPD, completeness, and accuracy of the recovered sample. For F090W and F115W dropouts, using redder filters in the UV continuum color cut leads to an increase in accuracy at a lower TPD and completeness. In addition, we recommend using the long wavelength NIRCam filters to assist in rejecting interloper galaxies for all three dropout schemes.
    \item The usage of the $\chi^2_{\mathrm{opt}}$ statistic \citep{bouwens2015} for removing low-redshift interloper galaxies is only recommended for samples with deep observations in multiple filters at wavelengths shorter than the Lyman break for the dropout sample, where assuming $\chi^2_{\mathrm{opt}} < 5$ would help to reject outliers. 
    \item While it will be possible for NIRCam to detect brown dwarf stars out to 10+ kpc, they are only a significant source of contamination in F090W dropout selection, which can be alleviated with color cuts using NIRCam data at 3 - 5$\mu$m. 
\end{enumerate}

\acknowledgments 
ECL acknowledges support from the ERC Advanced Grant 695671 ``QUENCH.'' CCW acknowledges support from the National Science Foundation Astronomy and Astrophysics Fellowship grant AST-1701546.

\bibliographystyle{apj}
\bibliography{apj-jour,colorrefs}

\end{document}